\documentclass{jfm}
\usepackage{graphicx}
\usepackage{newtxtext}
\usepackage{newtxmath}
\usepackage{natbib}
\usepackage{hyperref}
\hypersetup{
    colorlinks = true,
    urlcolor   = blue,
    citecolor  = black,
}

\newcommand{\RomanNumeralCaps}[1]
\linenumbers
\usepackage{comment}
\usepackage{color}
\usepackage{subcaption}
\usepackage{bm}% bold math

\usepackage[OMLmathsfit,sfdefault=cmbr]{isomath}
\newcommand\mathtensor[1]{\mathsfbfit{#1}}

\newcommand{\MDG}[1]{{\textcolor{red}{*** #1 ***}}}

\newcommand{\Wi}{\textit{Wi}}
\newcommand{\Sc}{\textit{Sc}}

\shorttitle{Nested traveling wave structures in EIT}
\shortauthor{M. Kumar and M. D. Graham}

\title{Nested traveling wave structures in elastoinertial turbulence}

\author{Manish Kumar\aff{1}
 \and Michael D. Graham\aff{1}  \corresp{\email{mdgraham@wisc.edu}}}

\affiliation{\aff{1}Department of Chemical and Biological Engineering, University of Wisconsin-Madison, 1415 Engineering Dr, Madison, WI 53706, USA}

\begin{document}

\maketitle
\begin{abstract}
Elastoinertial turbulence (EIT) is a chaotic flow resulting from the interplay between inertia and viscoelasticity in wall-bounded shear flows. Understanding EIT is important because it is thought to set a limit on the effectiveness of turbulent drag reduction in polymer solutions. Here, we analyze simulations of two-dimensional EIT in channel flow using Spectral Proper Orthogonal Decomposition (SPOD), discovering a family of traveling wave structures that capture the sheetlike stress fluctuations that characterize EIT.  The frequency-dependence of the leading SPOD mode contains distinct peaks and the mode structures corresponding to these peaks exhibit well-defined traveling structures. The structure of the dominant traveling mode exhibits shift-reflect symmetry similar to the viscoelasticity-modified Tollmien–Schlichting (TS) wave, where the velocity fluctuation in the traveling mode is characterized by large-scale regular structures spanning the channel and the polymer stress field is characterized by thin, inclined sheets of high polymer stress localized at the critical layers near the channel walls. The traveling structures corresponding to the higher-frequency modes have a very similar structure, but are nested in a region roughly bounded by the critical layer positions of the next-lower frequency mode. A simple theory based on the idea that the critical layers of mode $\kappa$  form the ``walls" for the structure of mode $\kappa+1$ yields quantitative agreement with the observed wave speeds and critical layer positions, indicating self-similarity between the structures. The physical idea behind this theory is that the sheetlike localized stress fluctuations in the critical layer prevent velocity fluctuations from penetrating them.
%Further, the structures corresponding to the successive energy peaks exhibit nested structures, where the mode structure corresponding to a following peak is bounded by the layers of stress fluctuations from the previous peak.}  %The characteristic structures of the traveling mode both in velocity and stress fields progressively concentrate toward the centerline of the channel as their wave speed increases and hence exhibit a nested arrangement of traveling waves. %We also develop a low-dimensional model of elastoinertial turbulence and discover that the reconstruction of the velocity field requires relatively fewer SPOD modes compared to the polymer stress field. 
\end{abstract}

%\begin{keywords}
%Authors should not enter keywords on the manuscript, as these must be chosen by the author during the online submission process and will then be added during the typesetting process (see http://journals.cambridge.org/data/\linebreak[3]relatedlink/jfm-\linebreak[3]keywords.pdf for the full list)
%\end{keywords}

\section{Introduction}
%Chaotic dynamics emerge in many nonlinear systems such as inertial turbulence, active turbulence \citep{Alert2022}, elastic turbulence \citep{Kumar2022review}, and elastoinertial turbulence \citep{Dubief2023}. For practical applications, it is desirable to manipulate these chaotic systems. For example, it is desirable to suppress inertial turbulence to minimize energy losses and promote elastic turbulence to enhance small-scale mixing and transport. The very high dimensionality of these chaotic systems creates enormous challenges for understanding or designing control strategies to manipulate their chaotic behaviors. However, their dynamics are often organized around some well-defined structures known as exact coherent structures (ECS), which provide a highly reduced-order representation of chaotic systems \citep{Graham2021}. Periodic, quasi-periodic, and traveling wave solutions of their governing equations are examples of ECS and these structures provide an excellent framework for understanding, predicting, and manipulating dynamics in chaotic systems. 

Adding a tiny amount of high molecular weight polymer to a fluid dramatically reduces turbulent drag \citep{toms1949some}. Therefore, the polymer additives are used to reduce pumping costs in pipeline transport of crude oil and home heating and cooling systems, and to reduce fuel transfer time in airplane tank filling \citep{Brostow2008}. Polymer additives also have been envisioned for flood remediation and enhancement of the drainage capacity of sewer systems \citep{Kumar2023,Bouchenafa2021,Sellin1978}. Newtonian turbulent flow contains streamwise vortices close to walls, which dominate the near-wall momentum transport and thus the drag. During drag reduction, the polymer chains get stretched due to turbulence, leading to stress distributions that wrap around the streamwise vortices, weakening them to lead to lower turbulent drag \citep{Li2007,Kim:2007dq,Graham2021}. %polymer chains in turbulent flows get stretched, which weakens near-wall vortices and leads to a reduced turbulent drag \citep{Li2007,Graham2021}. 

However, this suppression of near-wall vortices does not generally lead to full relaminarization, but rather to a limiting state called the maximum drag reduction (MDR) asymptote. Some understanding of this observation has come from the discovery of elastoinertial turbulence (EIT), a complex chaotic flow that is sustained, rather than suppressed by viscoelasticity, and thus helps explain the absence of relaminarization \citep{Samanta2013,Dubief2023,Shekar:2019hq}. Nevertheless, the structure and mechanism underlying EIT remain poorly understood and are the topic of the present work. 

% The emergence of elastoinertial turbulence (EIT), a chaotic state resulting from the interplay between inertia and elasticity, is suspected behind the chaotic nature of MDR and hence suspected to set the limit of achievable drag reduction through polymer additives \citep{Samanta2013,Dubief2023}. %The emergence of elastoinertial turbulence is suspected to play a critical role in the MDR state \citep{Samanta2013,Dubief2023}. 

\begin{figure}
\centering
\includegraphics[width=\textwidth]{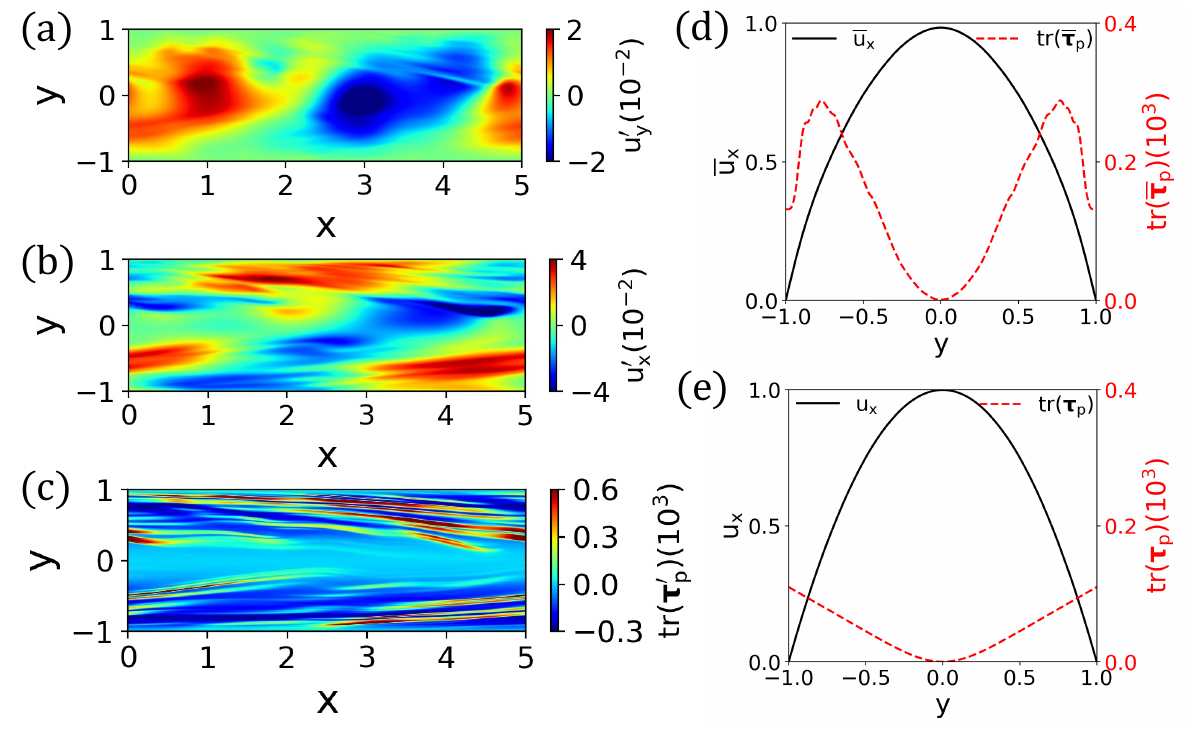}
\caption{Snapshots of perturbations of (a) wall-normal velocity  ($u_y'$), (b) streamwise velocity ($u_x'$), and (c) trace of the polymer stress tensor ($\mathrm{tr}(\mathtensor{\tau}_{p}^{\prime})$) from their temporal arithmetic means in EIT. (d) Profiles of mean streamwise velocity ($\overline{u}_x$) and mean of the trace of polymer stress tensor ($\mathrm{tr}(\overline{\mathtensor{\tau}}_{p})$) in EIT. (e) Profiles of streamwise velocity and the trace of the polymer stress tensor in the unidirectional laminar flow state. For all plots, $\Rey=3000$ and $\Wi=35$. \textcolor{black}{The variables have been nondimensionalized with their respective scales (see Section \ref{formulation_equations}).} }
\label{uy_thetaxx_prime_Wi35_EIT.pdf}
\end{figure}
EIT and MDR arise in parameter regimes of Reynolds number $\Rey$ and Weissenberg number $\Wi$ (product of polymer relaxation time and nominal strain rate) where the unidirectional laminar flow is linearly stable and is thus a nonlinearly self-sustaining flow.  The basic structure in both channel and pipe flows \citep{Samanta2013,Lopez:2019ct} is two-dimensional (2D) \citep{Sid:2018gh}, characterized by vorticity fluctuations localized in narrow regions near the walls with tilted sheets of highly stretched polymers emanating from these regions. \textcolor{black}{Figure \ref{uy_thetaxx_prime_Wi35_EIT.pdf} shows snapshots of velocity and polymeric stress fields of a simulated 2D EIT in channel flow along with their temporal mean profiles in EIT and the corresponding profiles in the unidirectional laminar state.} 

\begin{figure}
\centering
\includegraphics[width=\textwidth]{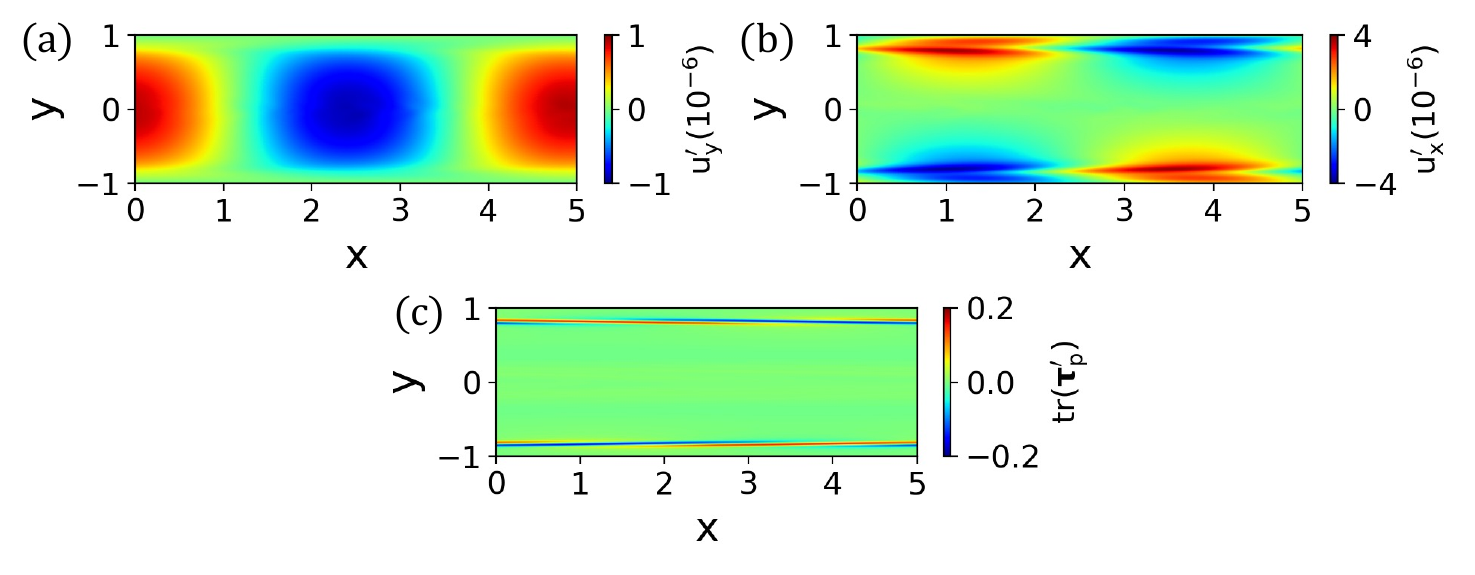}
\caption{Structures of the perturbations of (a) wall-normal velocity ($u_y'$), (b) streamwise velocity ($u_x'$), and (c) trace of polymer stress tensor ($\mathrm{tr}(\mathtensor{\tau}_{p}^{\prime})$) from the unidirectional laminar state for the viscoelastic linear TS wave at $Re=3000$ and $Wi=35$. This mode ultimately vanishes as viscoelastic channel flow is linearly stable at this parameter regime.}
\label{viscoelastic_linear_TS_Re3000_Wi35.pdf}
\end{figure}

Despite the absence of an obvious linear instability mechanism for EIT, it has been hypothesized that EIT is related to the nonlinear excitation of either a ``wall mode" or a ``center mode" structure arising in the linear stability problem for the unidirectional laminar state \citep{drazin1981hydrodynamic,Datta.2022.10.1103/physrevfluids.7.080701}.  A wall mode has a wave speed much less than the centerline velocity, and critical layer positions, i.e.~where the wave speed equals the local laminar velocity, near the walls. In contrast, a center mode travels at nearly the centerline velocity and accordingly the critical layer position is near the centerline.  The Tollmien-Schlichting (TS) mode of classical linear analysis of plane Poiseuille flow is a wall mode, and there is a strong structural resemblance of the viscoelastic extension of the TS wave to EIT \citep{Shekar:2019hq,Shekar.2020.10.1017/jfm.2020.372,Shekar.2021.10.1103/physrevfluids.6.093301qrb}. Figure \ref{viscoelastic_linear_TS_Re3000_Wi35.pdf} shows the viscoelastic linear TS mode at the same conditions as Figure \ref{uy_thetaxx_prime_Wi35_EIT.pdf}. \textcolor{black}{Since the laminar state at these conditions is linearly stable, a sufficiently small random perturbation will decay, with the slowest mode of decay having this form.} Sheets of highly stretched polymer are generated in the TS wave due to the presence of hyperbolic stagnation points (in the frame traveling with the wave) in the Kelvin cat's-eye structure of the velocity field in the critical layer \citep{Shekar:2019hq}. \textcolor{black}{The recently discovered ``polymer diffusive instability (PDI)" is also a wall mode \citep{Beneitez2023,Couchman2024}. However, in the parameter regime considered in the present study, the PDI does not arise; the laminar profile is linearly stable, and accordingly simulations with initial conditions that are very small perturbations from laminar flow decay back to it.} The possibility of a center mode structure is of interest in part because there is a linear center mode instability at low Reynolds number $\Rey$ that may organize ``elastic turbulence" at very small Reynolds number $\Rey$ \citep{Garg2018,Khalid2021,Choueiri.2021.10.1073/pnas.2102350118,Morozov.2022.10.1103/physrevlett.129.017801}.  
Nevertheless, in the elastoinertial regime considered here, while center mode structures can exist \citep{Dubief2022}, they do not appear to play an active role in the structure and self-sustenance of EIT \citep{Beneitez2024}.    
 The present work is consistent with this picture, and indeed deepens the connection between EIT and wall modes, showing in particular the existence of a nested family of such structures.

In the present study, we investigate the structure and dynamics of EIT in channel flow using a modal decomposition technique known as spectral proper orthogonal decomposition (SPOD) \citep{Towne2018}. SPOD characterizes coherent structures in complex flows that have well-defined structures and persist both in space and time via a frequency-domain variant of Proper Orthogonal Decomposition (POD) \citep{lumley1967structure}. At a given frequency, SPOD generates an energetically ordered and spatially orthogonal set of modes characterizing the flow.
% where the SPOD analysis requires a time-series of spatially resolved chaotic dynamics and it generates oscillating modes at each frequency which are spatially orthogonal to each other. 
The modes and eigenvalues obtained using SPOD analysis can be interpreted as physical structures with a particular frequency and the energies associated with those structures \citep{Schmidt2019}. \textcolor{black}{SPOD and dynamic mode decomposition (DMD) are related, as detailed in \cite{Towne2018}. In particular, DMD gives modes that correspond to particular frequencies. However, in DMD, there is no natural ordering of modes; in contrast, SPOD gives at each frequency an orthogonal basis set of modes ordered by their mean-square contribution to the flow at that frequency.}

SPOD has been successfully used in inertial turbulence to understand coherent structures and develop a low-dimensional model for turbulence \citep{Schmidt2017,Araya2017,Tutkun2017,Braud2004,Hellstrom2014,Nekkanti2021}. Here, we use it to investigate traveling coherent structures underlying the chaotic dynamics of EIT.  We will focus on the frequency-dependence of the most energetic SPOD mode, as it reveals important coherent features of the flow. 
% \textcolor{black}{Dynamic Mode Decomposition (DMD) also gives coherent structures that persist in both space and time and is related to SPOD as the SPOD modes are optimally averaged DMD modes \citep{Towne2018}. However, we prefer to use SPOD in the present study as it optimally accounts for the statistical variability of the turbulent flow.} 

% and also develop a low-dimensional model of EIT using the SPOD modes.

\section{Formulation and governing equations}\label{formulation_equations}
Because the self-sustaining dynamics of elastoinertial turbulence are fundamentally 2D \citep{Sid:2018gh}, we consider two-dimensional (2D) viscoelastic channel flow with nondimensional equations of mass and momentum conservation:
\begin{equation}\label{colm}
\nabla \cdot \boldsymbol{u}=0, \ \ \ \ \ \ \frac{\partial \boldsymbol{u}}{\partial t}+\boldsymbol{u}\cdot \nabla \boldsymbol{u} =-\nabla p+\frac{\beta}\Rey \nabla^2 \boldsymbol{u} + \frac{1-\beta}\Rey \nabla \cdot \mathtensor{\tau}_p+f(t)\boldsymbol{e}_x,
\end{equation}
where $\boldsymbol{u}$ and $p$ are non-dimensional velocity field and pressure field, respectively. Newtonian laminar centerline velocity ($U_c$) and channel half width ($H$) have been used as characteristic velocity scale and length scale, respectively. The ratio between solvent viscosity ($\eta_s$) to zero shear rate solution viscosity ($\eta$) has been denoted by $\beta=\eta_s/\eta$. The Reynolds number has been defined as $\Rey=\rho U_c H/\eta$, where $\rho$ represents fluid density. We use no-slip boundary conditions for the velocity field at the channel wall. Periodic boundary conditions have been used at the inlet and outlet of the channel. Flow is driven by an external forcing term $f(t)\boldsymbol{e}_x$,  where $\boldsymbol{e}_x$ denotes the streamwise direction. The forcing term varies with time to keep the \textcolor{black}{bulk velocity ($U_b=2U_c/3$) at its Newtonian laminar value}.
% enforce a constant volumetric flow rate:
% \begin{equation}\label{flow_rate}
% \int_{-1}^1 u_x dy=4/3.
% \end{equation}
 The polymer stress tensor is denoted  $\mathtensor{\tau}_p$ and we choose the FENE-P constitutive model with an artificial diffusion term to model its evolution: 
\begin{equation}\label{tau_theta}
\frac{\partial \mathtensor{\alpha}}{\partial t}+\boldsymbol{u}\cdot \nabla \mathtensor{\alpha} - \mathtensor{\alpha}\cdot \nabla \boldsymbol{u}-(\mathtensor{\alpha}\cdot \nabla \boldsymbol{u})^T=-\mathtensor{\tau}_p+\frac{1}{\Rey \Sc} \nabla^2 \mathtensor{\alpha},
\end{equation}
\begin{equation}\label{fenep}
\mathtensor{\tau}_p= \frac{1}{\Wi} \left( \frac{\mathtensor{\alpha}}{1-\mathrm{tr}(\mathtensor{\alpha})/b}-\mathtensor{I}\right),
\end{equation}
where $\mathtensor{\alpha}$ is the conformation tensor and parameter $b$ characterizes the maximum extensibility of the polymer chains. \textcolor{black}{The polymer stress tensor is nondimensionalized with  $\eta U_c/H$.} The Weissenberg number $\Wi=\lambda U_c/H$, where $\lambda$ is the polymer relaxation time. The Schmidt number $Sc=\eta/\rho D$, where $D$ is the diffusion coefficient, represents the ratio of momentum diffusivity to mass diffusivity. 
% The evolution equation of the conformation tensor is hyperbolic (Eq. \ref{tau_theta}).
% Therefore, 
The artificial diffusion term is used to stabilize the numerical scheme during the integration of Eq. \ref{tau_theta}. The presence of this term leads to the requirement of boundary conditions for the conformation tensor. At the channel walls, we determine $\mathtensor{\alpha}$ by solving the governing equations considering $Sc \to \infty$.

We solve the governing equations with a spectral method using the Dedalus framework \citep{Burns2020}. The computational domain has a length $L=5$.  The governing equations have been discretized using 256 Fourier basis functions and 1024 Chebyshev basis functions in the streamwise ($x$) and wall-normal ($y$) directions, respectively. We use $\beta=0.97$, $b=6400$, $\Rey=3000-6000$, and $\Wi=35-70$, which are relevant to turbulent drag reduction and are in the range of values used in previous studies, e.g.~\citep{Shekar:2019hq,Dubief2022}. In reality, the Schmidt number for a polymer solution is very large ($Sc\sim 10^6$). Numerical simulation with such a large value of $Sc$ would require an extremely fine mesh and small time-step, making numerical simulations computationally very expensive. At the same time, small $Sc$ (i.e., $Sc<10$) smears out small-scale dynamics and suppresses EIT \citep{Sid:2018gh,Dubief2023}. Therefore, numerical simulations of EIT based on artificial diffusion generally use $\Sc\sim O(100)$ \citep{Sid:2018gh,Buza2022b}. In the present study, we use $\Sc=250$ which is sufficient to sustain EIT and also numerically tractable. Viscoelastic channel flow in the parameter regime considered here is linearly stable. \textcolor{black}{Therefore, to trigger EIT, we use unidirectional laminar flow with sufficiently large random perturbations in the conformation tensor as the initial condition of the simulation.} In computing statistics, initial transients are dropped so that we consider only statistically stationary results.  %Then we use this EIT as the initial condition to simulate EIT in different parameter regimes.  

%\MDG{Check JFM style guide -- we may have to use different fonts for tensors/matrices}
\textcolor{black}{To estimate the SPOD spectrum, the spatiotemporal state variables are organized in a vector $\boldsymbol{q}(\boldsymbol{x},t)$. Here, we separately take this vector to contain wall-normal velocity, streamwise velocity, or the trace of the polymer stress tensor, as further discussed below. For a statistically stationary flow, the SPOD analysis can be done in Fourier space, where $\Tilde{\boldsymbol{q}}(\boldsymbol{x},f)$ denotes the Fourier-transformed dataset. Notation and methodology here follow \cite{Towne2018}. Given an inner product
\begin{equation}\label{inner_product}
\langle\boldsymbol{\Tilde{q},\psi}\rangle= \int_{\Omega}   \boldsymbol{\psi}^*(\boldsymbol{x},f) \Tilde{\boldsymbol{q}}(\boldsymbol{x},f) d\boldsymbol{x} 
\end{equation}
where $(^*)$ denotes conjugate transpose, the SPOD seeks to find a function $\boldsymbol{\psi}(\boldsymbol{x},f)$ that maximizes
\begin{equation}\label{objective_fn}
E\{ | \langle \Tilde{\boldsymbol{q}}(\boldsymbol{x},f), \boldsymbol{\psi}(\boldsymbol{x},f) \rangle |^2\}
\end{equation}
given the constraint $\langle \boldsymbol{\psi}(\boldsymbol{x},f), \boldsymbol{\psi}(\boldsymbol{x},f) \rangle=1$, where $\boldsymbol{\psi}(\boldsymbol{x},f)$ is the SPOD mode at frequency $f$. The operators $E\{\cdot\}$ and $|\cdot|$ represent expectation and modulus, respectively. 
% In the present study, we calculate the SPOD spectra of velocity and stress fields separately \citep{Wang2014}.
The maximization of Eq. \ref{objective_fn} leads to the following eigenvalue problem:
\begin{equation}\label{eigen_continous}
\int_{\Omega} \mathtensor{S}(\boldsymbol{x,y},f)  \boldsymbol{\psi}(\boldsymbol{y},f)  d\boldsymbol{y}=\sigma (f)  \boldsymbol{\psi}(\boldsymbol{x},f),
\end{equation}
where the cross-spectral density tensor $\mathtensor{S}$ is
\begin{equation}\label{cross_density_continous}
\mathtensor{S}(\boldsymbol{x,y},f) =E\{\Tilde{\boldsymbol{q}}(\boldsymbol{x},f) \Tilde{\boldsymbol{q}}^*(\boldsymbol{y},f) \}.
\end{equation}
% The cross-spectral density tensor is compact. 
This eigenvalue problem (Eq. \ref{eigen_continous}) leads to an infinite set of eigenmodes $\{\sigma_j (f), \boldsymbol{\psi}_j(\boldsymbol{x},f)\}$ at each frequency, which are generally arranged in decreasing order of $\sigma_j$. These eigenvalues are often called ``energies", because in the case where the time series is velocity, each eigenvalue indicates the amount of kinetic energy associated with the associated mode. The eigenvectors ($\boldsymbol{\psi}_j$) are orthogonal and provide a complete basis for $\Tilde{\boldsymbol{q}}$. Therefore, the Fourier-transformed dataset at a given frequency can be written in terms of SPOD modes as 
\begin{equation}\label{fourier_mode}
\Tilde{\boldsymbol{q}}(\boldsymbol{x},f)=\sum_{j=1}^{\infty}a_j(f)\boldsymbol{\psi}_j(\boldsymbol{x},f),
\end{equation}
where $a_j(f)=\langle \Tilde{\boldsymbol{q}}(\boldsymbol{x},f), \boldsymbol{\psi}_j(\boldsymbol{x},f) \rangle$ are the SPOD coefficients.} % \MDG{Above paragraph needs a rewrite. In \eqref{fourier_mode}, $\boldsymbol{\psi}$ needs to be replaced by $\boldsymbol{\psi}_j$, but you haven't introduced the eigenvalue problem for the modes yet so you can't define $\boldsymbol{\psi}_j$.    }

\textcolor{black}{To calculate the SPOD of a discrete time series of $N_t$ snapshots $\left\{\boldsymbol{q}(t_1),\boldsymbol{q}(t_2),\ldots,\boldsymbol{q}(t_{N_t})\right\}$, first, a data matrix $\mathtensor{Q}$ is constructed as:
\begin{equation}\label{matrix_Q}
\mathtensor{Q}=[\boldsymbol{q}_1, \boldsymbol{q}_2, ...., \boldsymbol{q}_{N_t}],
\end{equation}
where $\boldsymbol{q}_i=\boldsymbol{q}(t_i)$. 
%\MDG{I have commented out the statement about SPOD not converging as the number of snapshots increase, because it's beside the point -- if you only have one time series, you need to split it into an ensemble of shorter ones regardless}
% The SPOD spectrum estimated directly using the data matrix $\boldsymbol{Q}$ does not converge as the number of snapshots increases \citep{bendat2000random}. Therefore, m
Multiple realizations of the flow field are generated by dividing the data matrix into overlapping blocks \citep{Welch1967}
\begin{equation}\label{matrix_Qn}
\mathtensor{Q}^n=[\boldsymbol{q}_1^n, \boldsymbol{q}_2^n, ..,\boldsymbol{q}_m^n,.., \boldsymbol{q}_{N_f}^n], \ n=1,2,..., N_b,
\end{equation}
where $N_f$ is the number of snapshots in each block. The total number of blocks can be given as $N_b=(N_t-N_o)/(N_f-N_o)$, where $N_o$ represents the number of overlapping snapshots. The $m^{th}$ entry in the $n^{th}$ block ($\boldsymbol{q}_m^n$) can be connected with the entry in $\mathtensor{Q}$ as $\boldsymbol{q}_m^n=\boldsymbol{q}_{m+(n-1)(N_f-N_o)}$. The non-periodicity of the data in each block may lead to spectral leakage during the estimation of the discrete Fourier transform (DFT). Therefore, to reduce the spectral leakage we compute the DFT of the windowed data:    
\begin{equation}\label{matrix_Qnw}
\mathtensor{Q}^{n,w}=[w_1\boldsymbol{q}_1^n, w_2\boldsymbol{q}_2^n, ..,w_m\boldsymbol{q}_m^n,.., w_{N_f}\boldsymbol{q}_{N_f}^n],
\end{equation}
where $w_m$ is the nodal value of the symmetric Hamming window function:
\begin{equation}\label{hamming}
w_m= 0.54-0.46\cos \left(\frac{2\pi (m-1)}{N_f-1} \right).
\end{equation}
The discrete Fourier transform of $\mathtensor{Q}^{n,w}$ gives
\begin{equation}\label{matrix_Qn_f}
\Tilde{\mathtensor{Q}}^n=[\Tilde{\boldsymbol{q}}_1^n, \Tilde{\boldsymbol{q}}_2^n, ..,\Tilde{\boldsymbol{q}}_m^n,.., \Tilde{\boldsymbol{q}}_{N_f}^n],
\end{equation}
where $\Tilde{\boldsymbol{q}}_m^n$ represents the Fourier component at frequency $f_m$ in the $n^{th}$ block. Next, the data matrix is organized frequency-wise, where the Fourier components at frequency $f_m$ from all the blocks are collected as
\begin{equation}\label{matrix_Qm_f}
\Tilde{\mathtensor{Q}}_m=[\Tilde{\boldsymbol{q}}_m^1, \Tilde{\boldsymbol{q}}_m^2, ...., \Tilde{\boldsymbol{q}}_{m}^{N_b}].
\end{equation}
Now, the SPOD modes, $\boldsymbol{\psi}_m$, and energies, $\mathtensor{\sigma}_m$, at the frequency $f_m$ can be obtained by computing the eigenvectors and eigenvalues of the discretized cross-spectral density (CSD) matrix $\mathtensor{S}_m=\Tilde{\mathtensor{Q}}_m\Tilde{\mathtensor{Q}}_m^*$ by solving the eigenvalue problem
\begin{equation}\label{eigenvalue_discrete}
 \mathtensor{S}_m \mathtensor{W} \boldsymbol{\psi}_m = \boldsymbol{\psi}_m \mathtensor{\sigma}_m,
\end{equation}
where $\mathtensor{W}$ is a positive-definite weighting matrix, which properly accounts for the numerical quadrature for integration on a non-uniform discrete grid, and $\mathtensor{\sigma}_m$ is a diagonal matrix of eigenvalues.  This equation is the discretized version of \eqref{eigen_continous}. In practice, the number of flow realizations ($N_b$) is much smaller than the number of grid points. Therefore, for faster computation, the eigenvalue problem 
\begin{equation}\label{eigenvalue_analogous}
\Tilde{\mathtensor{Q}}_m^* \mathtensor{W} \Tilde{\mathtensor{Q}}_m \boldsymbol{\Theta}_m = \boldsymbol{\Theta}_m \mathtensor{\sigma}_m
\end{equation}
is solved. This has the same nonzero eigenvalues as \eqref{eigenvalue_discrete}, and its eigenvectors are related to those of \eqref{eigenvalue_discrete} by the expression 
\begin{equation}\label{eigenvector}
\boldsymbol{\psi}_m = \Tilde{\mathtensor{Q}}_m\boldsymbol{\Theta}_m \mathtensor{\sigma}_m^{-1/2}.
\end{equation}
}

\begin{comment}
where $\boldsymbol{W}$ is a positive-definite weighting tensor, which properly weights the contribution of different state variables to the total energy \citep{Wang2014} and also accounts for the numerical quadrature for integration on a non-uniform discrete grid. In the present study, we calculate the SPOD spectra of velocity and stress fields separately. Therefore, $\boldsymbol{W}$ only needs to account for the integration on the non-uniform grid and it can be given as
\begin{equation}\label{weight}
\boldsymbol{W}= \int_{\Omega}  d\boldsymbol{x}.
\end{equation}

\MDG{Something's wrong here -- this is a constant, so it cannot account for a nonuniform grid. And in any case, you have written this for the continuum field $q(x)$ -- issues with grid don't arise until you discretize. Rewrite to be more precise}  
\end{comment}

For the SPOD analysis, we use the MATLAB tool developed by \cite{Schmidt2022}. Details of the method and its numerical implementation can be found in literature \citep{Towne2018,Schmidt2020}. We use $600$ time units of data generated using EIT simulation to perform SPOD analysis, which is sufficient for the convergence of SPOD. The dataset consists of $N_t=8000$ snapshots, which are sampled at the interval of $\Delta t_s=0.075$ time units. %To obtain converged spectral densities, multiple realizations of the flow are used \citep{bendat2000random}. This can be achieved by dividing the data set into multiple overlapping blocks \citep{Welch1967}. 
In the present study, we use $N_f=500$ snapshots in each block with $50 \%$ overlap ($N_o=250$), which leads to a total of $N_b=31$ blocks. The SPOD spectra estimated using different combinations of $N_f$ and $N_o$ have been shown in Appendix \ref{spod_Nf_No}. The number of modes obtained in SPOD is the same as the number of blocks, where the first mode has the highest energy and the last mode has the lowest energy. The number of non-negative frequencies is given by $N_f/2+1$ and the interval between discrete consecutive frequencies is $\Delta f= 1/ (\Delta t_s N_f)$. 

\section{Results and discussion}

This study focuses on the case $\Rey=3000, \Wi=35$, for which a snapshot of perturbation of wall-normal velocity $u_y'$, streamwise velocity $u_x'$, and the trace of the polymer stress tensor $\mathrm{tr}(\mathtensor{\tau}_p')$ are shown in Figure \ref{uy_thetaxx_prime_Wi35_EIT.pdf}, where $(^{\prime})$ represents the perturbation from the temporal arithmetic mean. \textcolor{black}{We have also plotted mean and laminar profiles of different state variables in Figure \ref{uy_thetaxx_prime_Wi35_EIT.pdf}. Note that the trace of the polymer stress field is closely related to the degree of polymer stretching, which is proportional to $\mathrm{tr}(\mathtensor{\alpha})$. The dynamics of $u_y'$ in EIT are dominated by the downstream advection of large-scale structures spanning the channel (Fig. \ref{uy_thetaxx_prime_Wi35_EIT.pdf}a and Supplementary video: Movie1) and the dynamics of $u_x'$ are dominated by the downstream advection of structures localized close to the channel walls (Fig. \ref{uy_thetaxx_prime_Wi35_EIT.pdf}b and Supplementary video: Movie2). The dynamics of the polymer stress field are dominated by the downstream motion of thin inclined sheets of polymer stress in the vicinity of the channel walls (Fig. \ref{uy_thetaxx_prime_Wi35_EIT.pdf}c and Supplementary video: Movie3).} Since wall-normal velocity is identically zero in the laminar state, its temporal mean is also identically zero so it yields the cleanest Fourier and SPOD spectra.  We note that the velocity fluctuations at EIT are generally quite small (e.g.~$u_y'\sim 10^{-2}$ and drag is only $12.8\%$ higher than the laminar), so the streamwise velocity profile does not greatly differ from laminar and hence it remains the dominant component of velocity (Fig. \ref{uy_thetaxx_prime_Wi35_EIT.pdf}). The trace of the polymer stress tensor represents the contribution of polymer chains in the stress field, which regulates the flow field. Therefore, $u_y'$, $u_x'$, $\mathrm{tr}(\mathtensor{\tau}_p')$ are the important variables to be analyzed using SPOD.

% \textcolor{black}{We consider $u_x^{\prime}$ to analyze the streamwise velocity component, where $(^{\prime})$ denotes deviation from the temporal arithmetic mean (Appendix \ref{mean_profile})\MDG{nope - this goes in the main text}. 

\textcolor{black}{We perform SPOD analyses of the perturbation fields $u_y'$, $u_x'$, and $\mathrm{tr}(\mathtensor{\tau}_{p}^{\prime})$ separately due to the intense memory requirements of the algorithm of \cite{Schmidt2022}. However, for a small dataset ($N_t=4000, N_o=125$), we have also calculated the SPOD spectrum of the velocity components together, which shows that the main characteristics of SPOD spectra remain unchanged (Appendix \ref{SPOD_together}). As will be shown, the spectral characteristics arising from each separate analysis are highly consistent, displaying peaks at the same frequencies. It is possible in principle, though highly memory intensive, to perform SPOD on the entire velocity and stress (or conformation) field, by extending the framework presented for POD by \cite{Wang2014}. As indicated there, as well as in \cite{Hameduddin2019feb}, some subtleties arise in working with tensors such as $\mathtensor{\alpha}$ that are constrained to be positive definite.}% Hence, our SPOD analyses of the polymer stress field will consider the deviation of the trace of polymer stress tensor from the mean ($\mathrm{tr}(\boldsymbol{\tau}_{p}^{\prime})$). It has been noted that for some statistics of conformation tensor (especially volume), geometric and log-Euclidean means are better suited compared to the arithmetic mean \citep{Hameduddin2019feb}. \MDG{this is confusing -- we're using stress, not stretch. What point are you trying to make here?} In the present study, stretch is the statistical quantity of interest for the conformation tensor. Therefore, we use the arithmetic mean and linear perturbations even for the polymer stress field, which also provides consistency with the SPOD analyses of the velocity field.} %\MDG{are we going to say anything about the Hamaduddin work?} \textcolor{green}{I have added a couple of sentences above discussing it.}%\MDG{Also we need to be clear on why we are analyzing these separately. Seems like we should at least be able to do the whole velocity field in one analysis.}

% Even though elastoinertial turbulence is a chaotic flow state, the velocity fluctuations in EIT are relatively weak (a few percent of centerline velocity). Therefore, we choose the wall-normal component of velocity ($u_y'$) to analyze the dynamics of EIT as this component is absent in laminar flow (Fig. \ref{uy_thetaxx_prime_Wi35_EIT.pdf}a). The chaotic nature of EIT is sustained by the fluctuation in the polymer stress field. To understand the coupling between the dynamics of flow velocity and polymer stress field, we also analyze the perturbation of the xx-component of the conformation tensor from the laminar state ($\alpha_{xx}^{\prime}$) as this component dominates the polymer stress field in EIT (Fig. \ref{uy_thetaxx_prime_Wi35_EIT.pdf}b). 

\begin{figure}
\centering
\begin{subfigure}[b]{0.49\textwidth} 
\includegraphics[width=\textwidth]{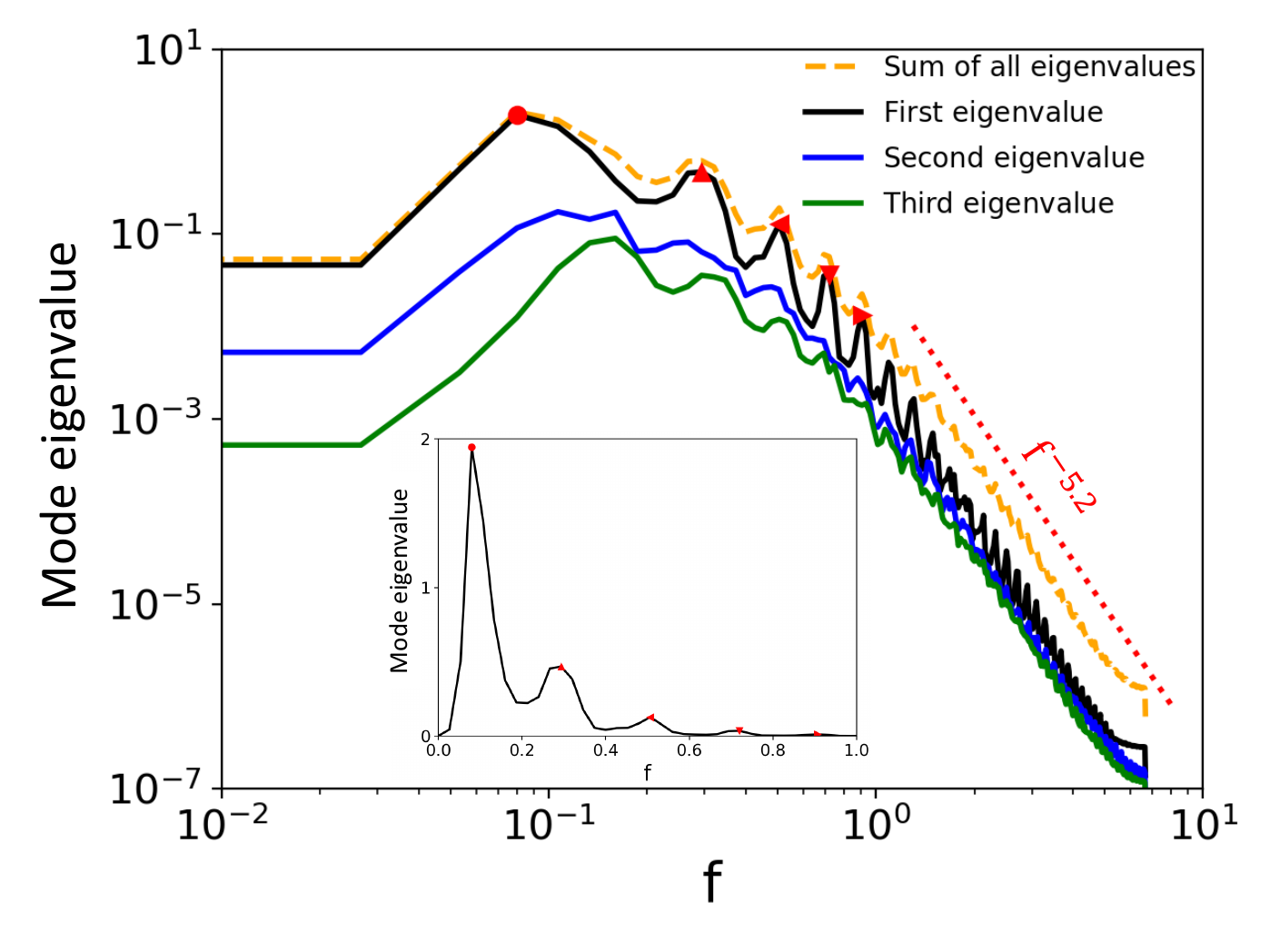}
\caption{}
\label{power_spectra_uy_Re3000_Wi35.pdf}
\end{subfigure}
\begin{subfigure}[b]{0.49\textwidth} 
\includegraphics[width=\textwidth]{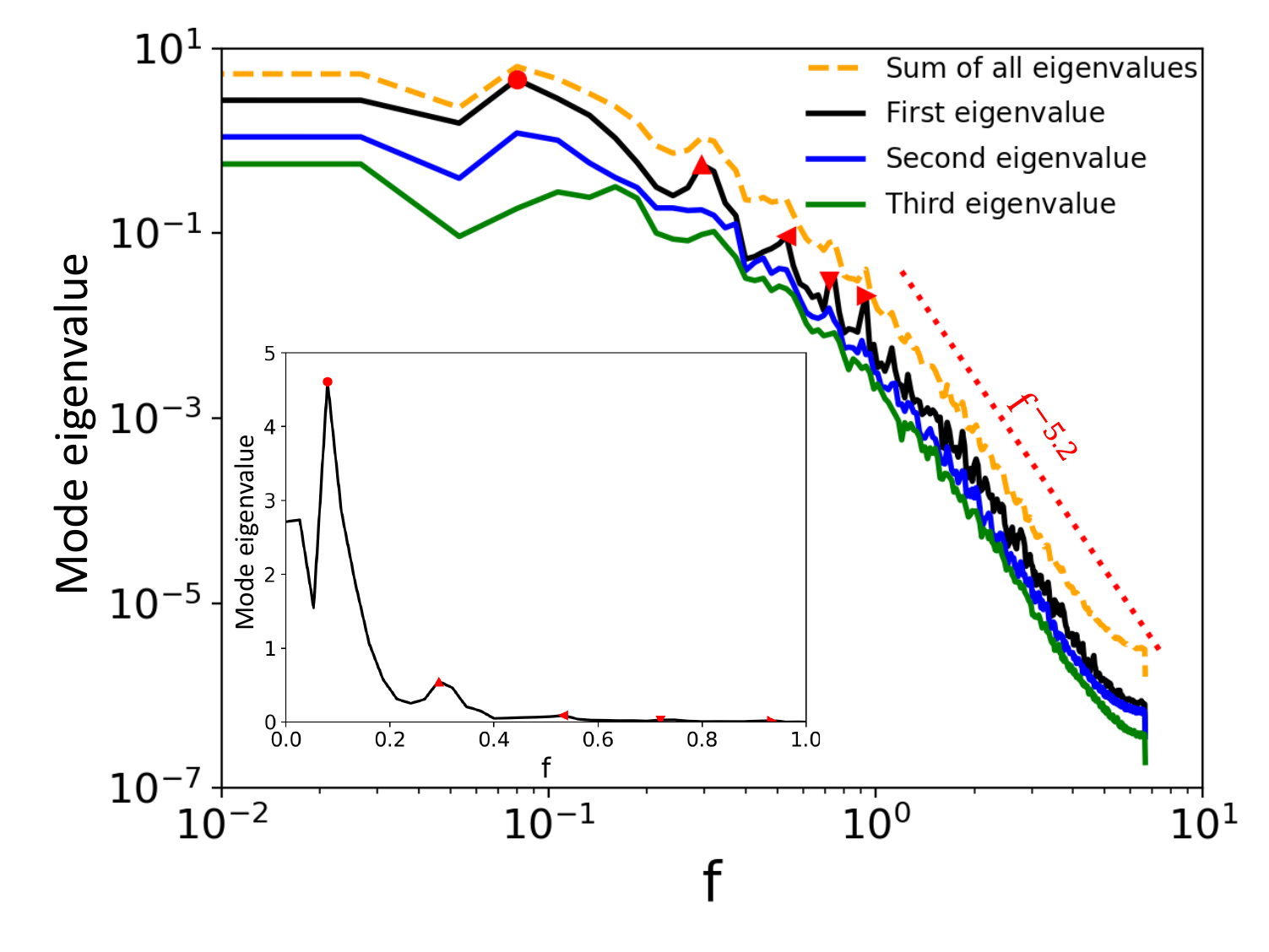}
\caption{}
\label{power_spectra_ux_Re3000_Wi35.pdf}
\end{subfigure}
\begin{subfigure}[b]{.49\textwidth}
\includegraphics[width=\textwidth]{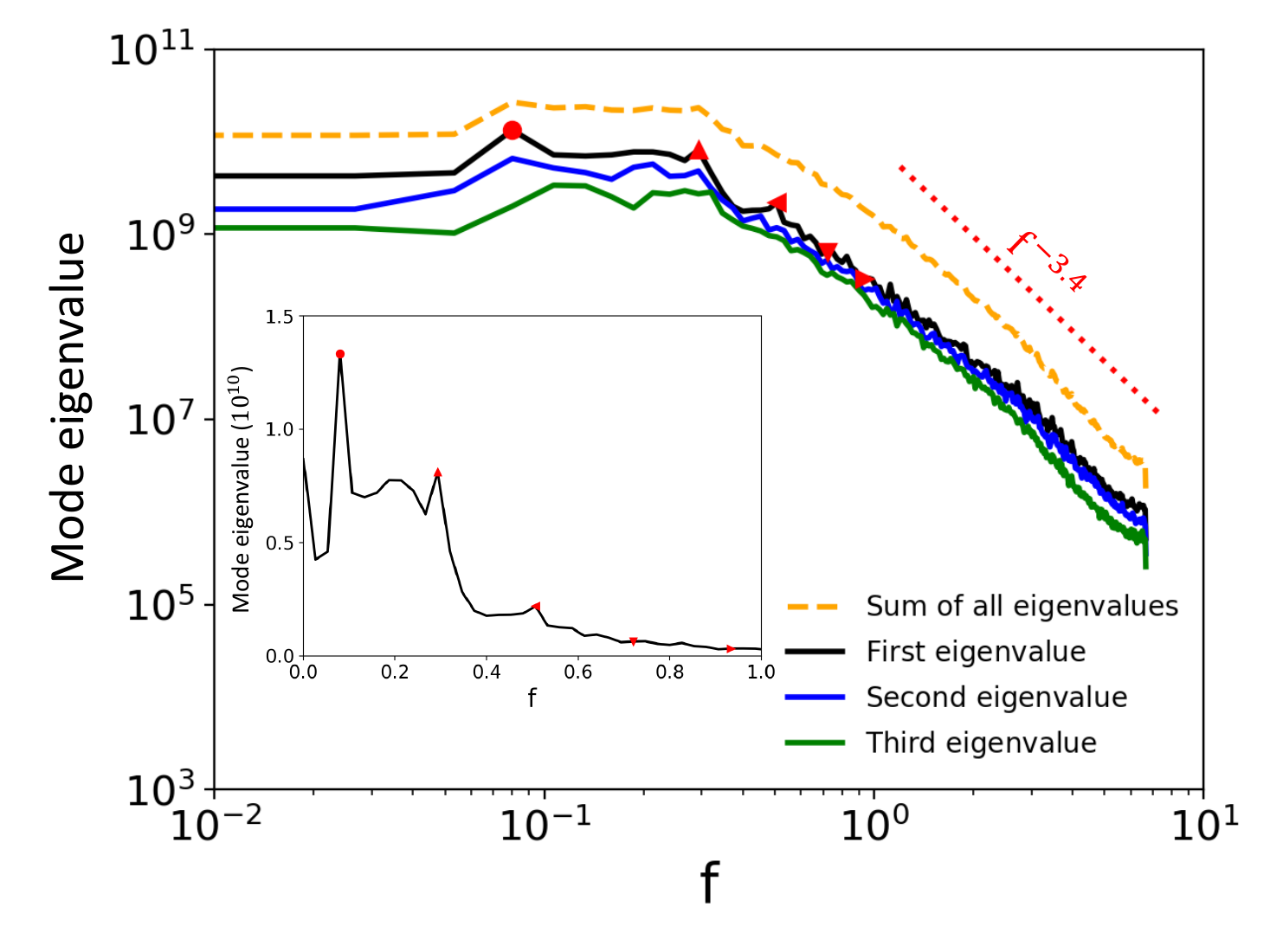}
\caption{}
\label{mode_energy_vs_frequency_thetaxx_Re3000_Wi35_nfft1k.png}
\end{subfigure}
\caption{SPOD eigenvalue spectra of perturbations of (a) wall-normal velocity ($u_y'$), (b) streamwise velocity ($u_x^{\prime}$), and (c) trace of polymer stress tensor ($\mathrm{tr}(\mathtensor{\tau}_{p}^{\prime})$) at $\Rey=3000$ and $\Wi=35$. Red symbols indicate the first few peaks in the leading mode of the eigenvalue spectra. Insets: SPOD eigenvalue spectra of the leading SPOD modes on a linear scale. }\label{power_specra_Wi35_Re3000_ux_uy_trtau}
\end{figure}

SPOD eigenvalue spectra as a function of frequency for the first several modes in $u_y'$, $u_x^{\prime}$, and $\mathrm{tr}(\mathtensor{\tau}_{p}^{\prime})$ are shown in Fig. \ref{power_spectra_uy_Re3000_Wi35.pdf}, Fig. \ref{power_spectra_ux_Re3000_Wi35.pdf}, and Fig. \ref{mode_energy_vs_frequency_thetaxx_Re3000_Wi35_nfft1k.png}, respectively,  along with the sum of eigenvalues of all the modes. \textcolor{black}{For velocity components, the eigenvalue represents the kinetic energy as mentioned earlier and the total kinetic energy can be represented by the SPOD amplitude as $||a||^2$ (Eq. \ref{fourier_mode}).} The leading modes in wall-normal velocity and streamwise velocity contain most of the energy ($\approx 73 \%$ and $\approx 55 \%$, respectively) and hence dominate the flow structure (Figs. \ref{power_spectra_uy_Re3000_Wi35.pdf} and \ref{power_spectra_ux_Re3000_Wi35.pdf}). In the eigenvalue spectrum of $\mathrm{tr}(\mathtensor{\tau}_{p}^{\prime})$, the leading mode has a relatively smaller contribution ($\approx 31 \%$) to $||a||^2$ (Fig. \ref{mode_energy_vs_frequency_thetaxx_Re3000_Wi35_nfft1k.png}). The leading modes of wall-normal velocity and streamwise velocity contain distinct sharp peaks at specific frequencies, of which the first few are indicated with red symbols, and the energy of these peaks decreases as the frequency increases. \textcolor{black}{The leading mode of the polymer stress field also has peaks at the same frequencies; they are not as sharp as the peaks for the velocity components but still quite distinct, as revealed by plotting on a linear scale (inset).} %The mode energy of the polymer stress field is several orders of magnitude ($\sim 10^{10}$) higher than that of velocity components.\MDG{You can't compare ``energy" between the stress and velocity here because you haven't done the SPOD in terms of a true energy norm -- leave this out} Therefore, even a tiny change in the relative energy close to the peaks in the SPOD spectrum of $tr(\boldsymbol{\tau}_p^{\prime})$ leads to a large difference in the absolute value. 
The higher-order modes do not have such distinct peaks. %\MDG{since you've put the power laws back in the figures, you need to say something about them in the text. Are these similar to what other people have seen? } \textcolor{green}{Done, Dubief 2013 reports a scaling -14/3 and Yamani 2021 reports -3. Our scaling is close to that of Dubief.} 
\textcolor{black}{The energy decay of velocity fluctuations in the SPOD spectra at a large frequency ($f>1$) approximately follows a ``power law" $f^{-5.2}$, which is somewhat close to the result $f^{-14/3}$ reported by \cite{Dubief2013}. The SPOD spectrum of $\mathrm{tr}(\mathtensor{\tau}_{p}^{\prime})$ follows a different power law ($f^{-3.4}$). The significance of these is unclear as they are only observed over less than a decade in frequency; we report them here only for completeness.}

\begin{figure}
\centering
\includegraphics[width=\textwidth]{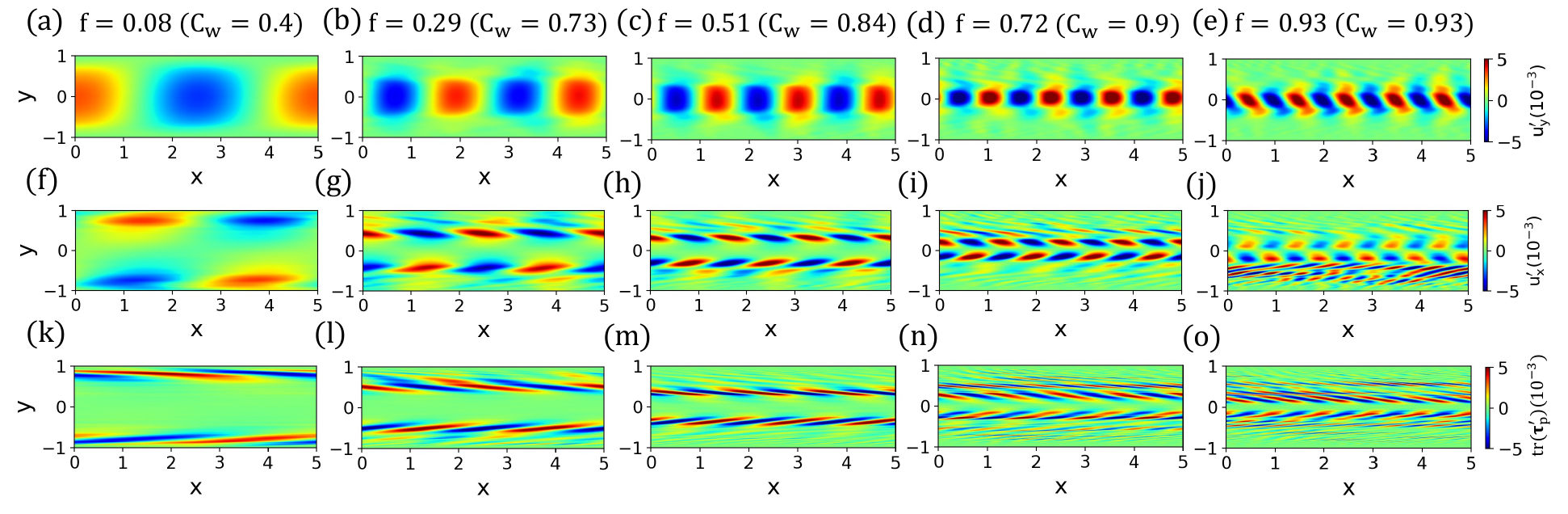}
\caption{Structures of SPOD modes of (a-e) $u_y'$, (f-j) $u_x^{\prime}$, and (k-o) $\mathrm{tr}(\mathtensor{\tau}_{p}^{\prime})$ at $\Rey=3000$ and $\Wi=35$; i.e., corresponding to the frequencies denoted by different symbols in the eigenvalue spectra (Figs. \ref{power_spectra_uy_Re3000_Wi35.pdf}, \ref{power_spectra_ux_Re3000_Wi35.pdf}, and \ref{mode_energy_vs_frequency_thetaxx_Re3000_Wi35_nfft1k.png}).}
\label{uy_different_frequency_Re3000_Wi35}
\end{figure}

The local peaks in the energy spectrum of velocity components indicate that the structures corresponding to these frequencies have distinct features in the dynamics of EIT. These peaks are \emph{not} at integer multiples of the lowest-frequency peak, so are not simply harmonics; the relationship between them is elucidated below.  The SPOD mode structures of $u_y'$, $u_x^{\prime}$, and $\mathrm{tr}(\mathtensor{\tau}_{p}^{\prime})$ corresponding to the peak frequencies in the leading SPOD mode have been shown in Fig. \ref{uy_different_frequency_Re3000_Wi35}. Each mode structure has a distinct wavenumber $\kappa$, which we measure in wavelengths per domain length.
These modes are all traveling waves with wave speed $C_w=fL/\kappa$, as further discussed below. 
% We will examine the structure of the dominant SPOD modes at the peak frequencies indicated with the red dots on Figure \ref{mode_energy_vs_frequency_thetaxx_Re3000_Wi35_nfft1k.png}.

The most dominant SPOD structure ($f=0.08$) has unit wavenumber ($\kappa=1$). The wall-normal velocity component for this mode consists of large-scale structures spanning the channel (Fig. \ref{uy_different_frequency_Re3000_Wi35}a), \textcolor{black}{the streamwise velocity component has regions of positive and negative velocity fluctuations close to the walls (Fig. \ref{uy_different_frequency_Re3000_Wi35}f)}, and the polymer stress field displays thin layers close to the walls having inclined alternating sheets of positive and negative stress fluctuations (Fig. \ref{uy_different_frequency_Re3000_Wi35}k). The structures approximately obey a shift-reflect symmetry: i.e.~$u_y(x,y)\approx -u_y(x+L/2, -y)$, $u_x^{\prime}(x,y)\approx u_x^{\prime}(x+L/2, -y)$, and $\mathrm{tr}(\mathtensor{\tau}_{p}^{\prime})(x,y) \approx \mathrm{tr}(\mathtensor{\tau}_{p}^{\prime})(x+L/2,-y)$. The quantification of the shift-reflect symmetry of different SPOD mode structures has been given in Appendix \ref{symmetry_quant}. This is the symmetry obeyed by the TS mode \citep{drazin1981hydrodynamic}, and comparison to Figure~\ref{viscoelastic_linear_TS_Re3000_Wi35.pdf} indicates a strong similarity in structure. 
% The sign of $u_y'$ in the adjacent large-scale structures in the velocity field alternates. Therefore, here onward, 
From here onward, we refer to the regions having positive and negative $u_y'$ as ``positive lobe" and ``negative lobe", respectively. The mode structures corresponding to other peaks have similar structures, where the wavenumber of structures increases with frequency (Fig. \ref{uy_different_frequency_Re3000_Wi35}). \textcolor{black}{The wall-normal extent of the lobes in $u_y'$ decreases and the regions of velocity fluctuations in $u_x^{\prime}$ approach the centerline of the channel as the wavenumber increases. Relatedly, the layers of strong $\mathrm{tr}(\mathtensor{\tau}_{p}^{\prime})$ move away from the wall as the frequency increases.}

\begin{figure}
\centering
\includegraphics[width=.8\textwidth]{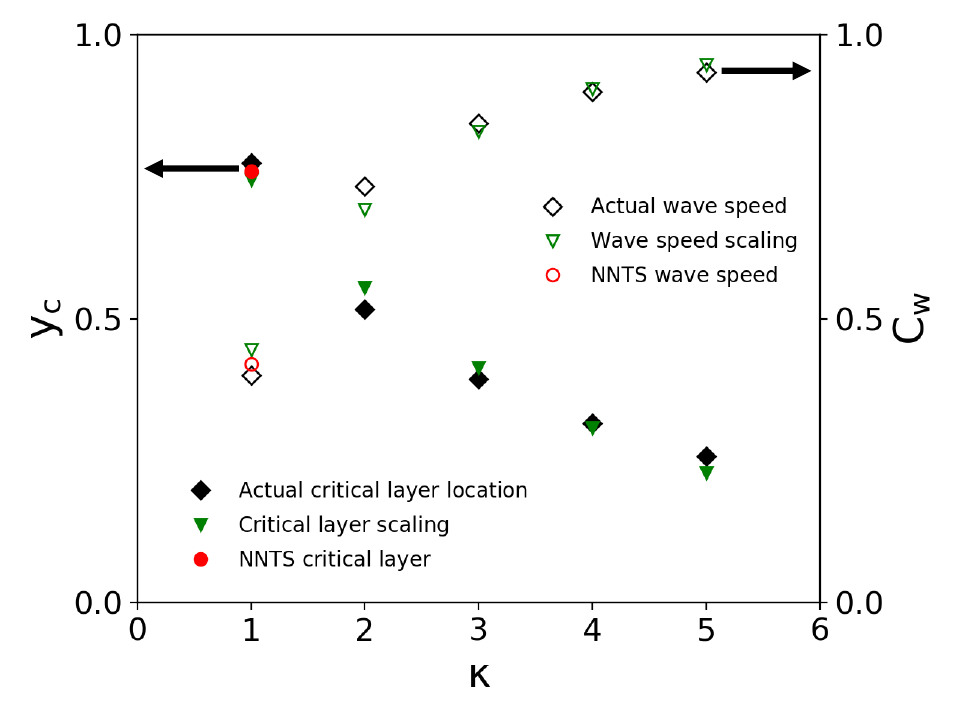}
\caption{Wave speed and location of the critical layer at $\Rey=3000$ and $\Wi=35$ for the traveling waves associated with the peaks in the leading SPOD mode along with the Newtonian nonlinear Tollmien–Schlichting (NNTS) wave results and the predictions of the scaling model Eqs.~\ref{eq:scalingy} and $\ref{eq:scalingc}$.}
\label{wave_speed_and_critical_layer_wave_number_nfft500.png}
\end{figure}

The wave speeds of the traveling structures in the leading SPOD mode are shown in Figure \ref{wave_speed_and_critical_layer_wave_number_nfft500.png}. They initially increase with the wavenumber (and frequency) and ultimately approach a value close to the centerline velocity of the channel ($C_w \to 0.94$). The wave speed of the dominant mode structure ($f=0.08,\kappa=1$) is very close to that of the Newtonian nonlinear TS (NNTS) wave at the corresponding $\Rey$ ($C_{TS}=0.42$), shown in red on Figure \ref{wave_speed_and_critical_layer_wave_number_nfft500.png}, further strengthening the evidence connecting EIT to the TS mode. \textcolor{black}{The NNTS wave belongs to the stable (in 2D) upper branch of the nonlinear traveling wave solution of plane Poiseuille flow, which originates at $\Rey=5772$ through a subcritical bifurcation from the laminar branch and exists down to $\Rey \approx 2800$  \citep{Jimenez1990,Shekar.2020.10.1017/jfm.2020.372}.} By contrast, a center mode would have a wave speed close to unity and thus a frequency close to $\kappa/L$. For $L=5$ this would be multiples of $0.2$, and Figure \ref{power_spectra_uy_Re3000_Wi35.pdf} shows no peaks at these positions. In fact, $f=0.2$ and its multiples are close to local minima in energy for the dominant SPOD mode. In short, we see no evidence of a center mode structure. The origin of the peak positions in Figure \ref{power_spectra_uy_Re3000_Wi35.pdf} is elucidated below.

\begin{figure}
\centering
\includegraphics[width=.6\textwidth]{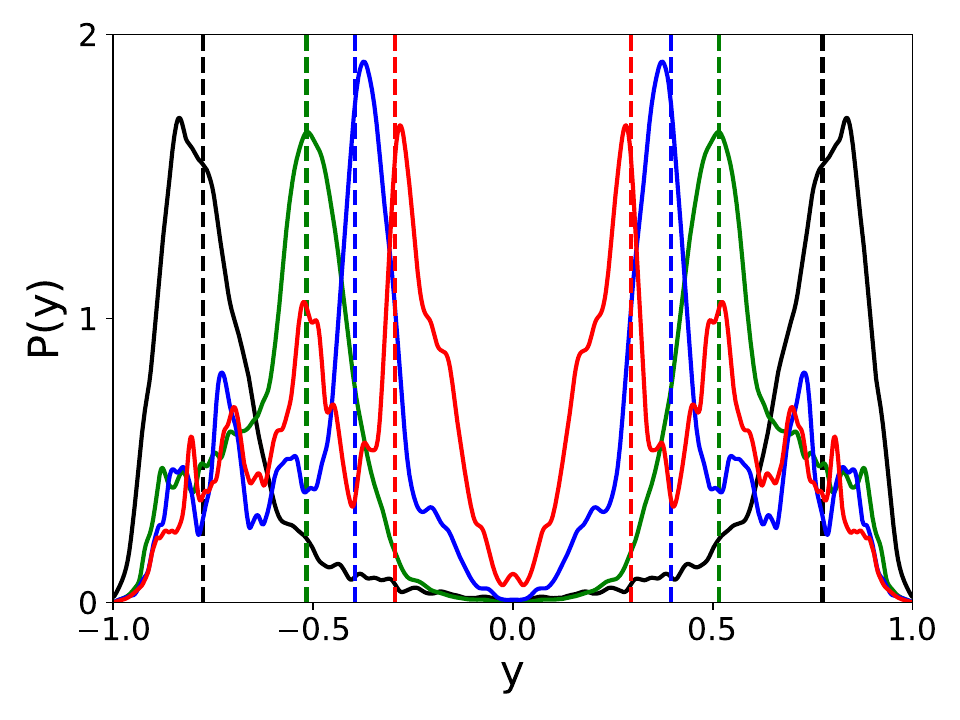}
\caption{Wall-normal distribution of polymer stress fluctuations $P(y)$ (solid lines) and the positions of critical layers (dashed lines) for traveling modes having wave speed $C_w=0.4$ (black), $C_w=0.73$ (green), $C_w=0.84$ (blue), and $C_w=0.9$ (red) at $\Rey=3000$ and $\Wi=35$.}
\label{thetaxx_xavg_vs_y}
\end{figure}

\begin{comment}
\begin{figure}[!ht]
\centering
\includegraphics[width=.5\textwidth]{thetaxx_xavg_vs_y_m0_f3_11_19_27_reflect_avg.png}
\caption{\label{thetaxx_xavg_vs_y} Streamwise $L_2$ norm of $\alpha_{xx}^{\prime}$ (solid lines) and critical layers (dashed lines) for traveling modes having wave speed $C_w=0.4$ (black), $C_w=0.73$ (green), $C_w=0.84$ (blue), and $C_w=0.9$ (red).}
\end{figure}
\end{comment}

As discussed in the Introduction, corresponding to the wave speed of a perturbation there is a critical layer position $y_c$. As this is the position where the fluid and the perturbation are moving together, it is the most favorable position for the two to exchange energy. 
% The concept of a ``critical layer" has been proven important to understand turbulent flows \citep{McKEON2010,Shekar:2019hq}. 
The velocity fluctuations in EIT are very weak so we can approximate the local streamwise velocity with the laminar value, $u_x\approx 1-y^2$, and the location of critical layers (i.e.~where $C_w=u_x$) can be given as $y_c\approx \pm \sqrt{1-C_w}$. Figure \ref{wave_speed_and_critical_layer_wave_number_nfft500.png} shows the critical layer positions corresponding to the wave speeds of the traveling structures of the leading SPOD mode, as well as for the NNTS mode. As with wave speed, the critical layer position of the $f=0.08, \kappa=1$ SPOD mode is very close to that of the NNTS mode.

\textcolor{black}{To illustrate the relation between the critical layer and the location of the peaks in the SPOD stress fluctuation structures, in Figure \ref{thetaxx_xavg_vs_y} we plot the positions of the critical layers for the traveling structures of the leading SPOD mode at the first several peak frequencies, along with the wall-normal distribution of polymer stress fluctuations ($P(y)$), which has been defined as 
\begin{equation}
P(y)=\left[ \frac{\int_0^L \{\mathrm{tr}(\mathtensor{\tau}_p^{\prime})\}^2 dx}{\int_{-1}^1 \int_0^L \{\mathrm{tr}(\mathtensor{\tau}_p^{\prime})\}^2 dx dy} \right]^{1/2}.\label{eq:Py}
\end{equation}}
The peak regions in $P(y)$ represent the locations of the sheets of polymer stress fluctuations, and we see that their locations correspond to the critical layers. A similar observation has been made for viscoelasticity-modified TS waves and it has been reported that thin sheets of high polymer stress emanate from the critical layers of TS waves \citep{Shekar:2019hq,Hameduddin2019}.

\begin{figure}
\centering
\includegraphics[width=\textwidth]{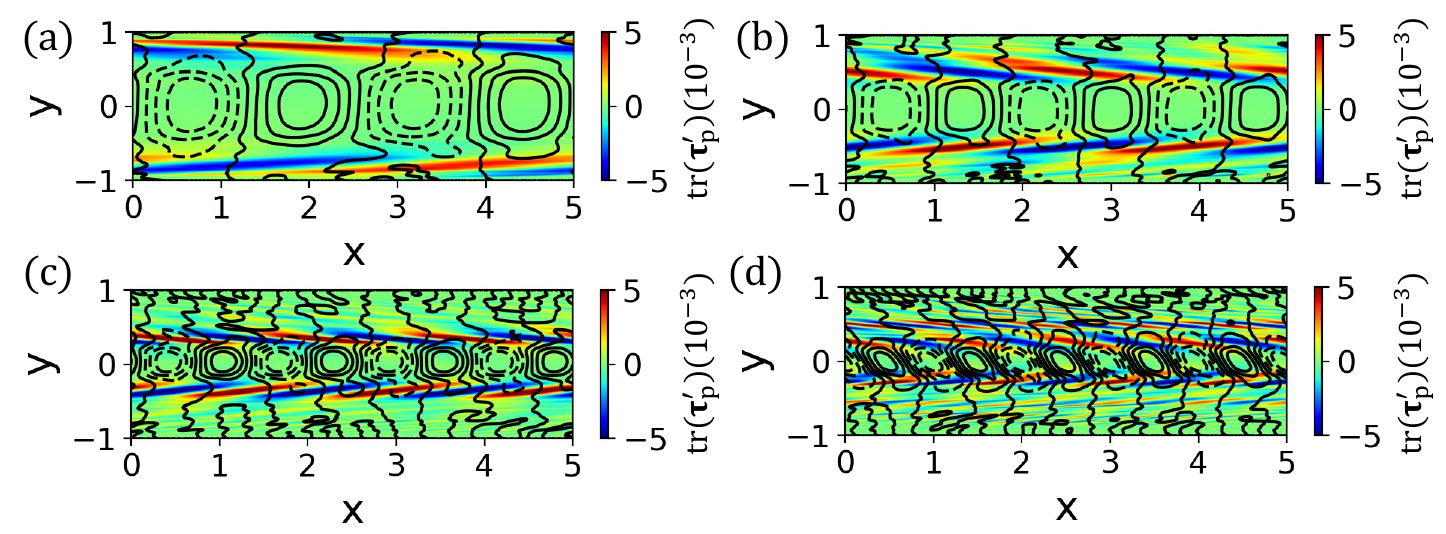}
\caption{Contours of velocity fluctuations ($u_y'$) of a faster-traveling wave on the top of the stress fluctuations of the immediate slower traveling wave: (a) stress at $C_w=0.4$ and velocity at $C_w=0.73$, (b) stress at $C_w=0.73$ and velocity at $C_w=0.84$, (c) stress at $C_w=0.84$ and velocity at $C_w=0.9$, and (d) stress at $C_w=0.9$ and velocity at $C_w=0.93$. Other parameters are $\Rey=3000$ and $\Wi=35$.}
\label{thetaxx_next_uy_contour_Re3000_Wi35.pdf}
\end{figure}

We noted above that as the wavenumber increases, $u_y'$ becomes more localized toward the channel center, as do the critical layer positions where the stress fluctuations are high. More specifically, consider the $u_y'$ profile at the second peak ($f=0.29$, Figure \ref{uy_different_frequency_Re3000_Wi35}b) and the $\mathrm{tr}(\mathtensor{\tau}_p^{\prime})$ profile at the first peak ($f=0.08$, Figure \ref{uy_different_frequency_Re3000_Wi35}k).  It appears that the ``lobes" where $u_y'$ is large in the former figure are roughly bounded by the layers where $\mathrm{tr}(\mathtensor{\tau}_p^{\prime})$ is large in the latter.  Similar observations can be made about all of the succeeding modes. We visualize this point in Figure \ref{thetaxx_next_uy_contour_Re3000_Wi35.pdf}, which replots the results of Figure \ref{uy_different_frequency_Re3000_Wi35} by showing contour lines of $u_y'$ from the SPOD modes at wavenumber $\kappa+1$ juxtaposed with color contours of $\mathrm{tr}(\mathtensor{\tau}_p^{\prime})$ at wavenumber $\kappa$.
From this figure, we see that the velocity lobes at wavenumber $\kappa+1$ are ``nested" within the stress fluctuations, or equivalently between the upper and lower critical layer positions, at wavenumber $\kappa$. 
In contrast, the regions between the critical layers and the channel walls contain small-scale and irregular structures both in the velocity field as well as stress field.

\textcolor{black}{We now present a simple theory for the results in Fig.~\ref{wave_speed_and_critical_layer_wave_number_nfft500.png} that is motivated by the above structural observations.}
The nested nature of the structures revealed by SPOD suggests that the locations of the polymer sheets of a slow-moving (low wavenumber) traveling wave act like ``walls" for the immediately faster-moving (and higher wavenumber) wave.  Consider the existence of a ``primary" mode with wave speed $C_{w,1}$ and thus critical layer positions $y_{c,1}=\pm(1-C_{w,1})^{1/2}$. We take the next higher mode to occupy the domain $|y|<| y_{c,1}|$; if its critical layer position $y_{c,2}$ is at the same fractional position in this new domain, it will thus be at $\pm |y_{c,1}|^2= \pm(1-C_{w,1})^{1}$. Continuing in this way, and noting that successively higher-speed waves can be labeled by their wavenumber $\kappa$, we have a simple scaling result
\begin{equation}
y_{c,\kappa}=(1-C_{w,1})^{\kappa/2}.\label{eq:scalingy}
\end{equation}
Relatedly, the successive wave speeds are then 
\begin{equation}
C_{w, \kappa}=1-(1-C_{w,1})^{\kappa}.\label{eq:scalingc}
\end{equation} 
Using the SPOD results for $y_{c,\kappa}$ to find a best-fit value of $C_{w,1}$ yields predictions for $y_{c,\kappa}$ and $C_{w,\kappa}$ that agree very closely with the data, as shown in Figure \ref{wave_speed_and_critical_layer_wave_number_nfft500.png}. Furthermore, the value of $C_{w,1}=0.44$ is very close to the NNTS wave speed $C_{TS}=0.42$. These observations indicate that the structure of EIT is dominated by nested self-similar structures that closely resemble TS waves.  
 
Finally, we briefly hypothesize a possible physical mechanism for the appearance of this nested structure. A highly stretched elastic sheet resists lateral deformation. Similarly, flows in which polymer molecules are strongly stretched along one direction resist deformations transverse to that direction. A classical example of this mechanism is the suppression of shear-layer instability in a viscoelastic fluid, where the strong stretching in the shear layer mimics an elastic sheet \citep{Azaiez:2006gs}. Relatedly, viscoelastic Taylor-Couette instability is suppressed by the normal stress induced by axial flow \citep{Graham:1998wa}, and in porous media flows, sheetlike regions with high polymer stress resist the flow passing through them and hence act like flow barriers \citep{Kumar2023stretching}. 

\textcolor{black}{A similar mechanism may be at work here, in which the sheets of high polymer stress in the critical layers from the primary mode prevent velocity fluctuations from the higher modes from passing through the critical layer, acting as ``walls" as noted above, and so on successively with the higher modes, leading to the emergence of a nested family of traveling waves. The resemblance of the SPOD mode structure in the reconstructed polymer stress field can be seen in Appendix \ref{trtaup_reconstruct}. 
% Thus, the formation of thin inclined sheets of high polymer stress at the critical layer of each traveling structure acts like ``walls" for the next-faster traveling structure and hence leads to the emergence of a nested family of traveling waves. 
Regardless of the detailed physical mechanism, the excellent agreement between the simulation results and the scaling theory manifested in Eqs.~\eqref{eq:scalingy}-\eqref{eq:scalingc} indicates the predictive power of the simple structural picture of nested traveling waves with critical layer fluctuations that define the length scale of the nesting.}

\begin{figure}
\centering
\includegraphics[width=\textwidth]{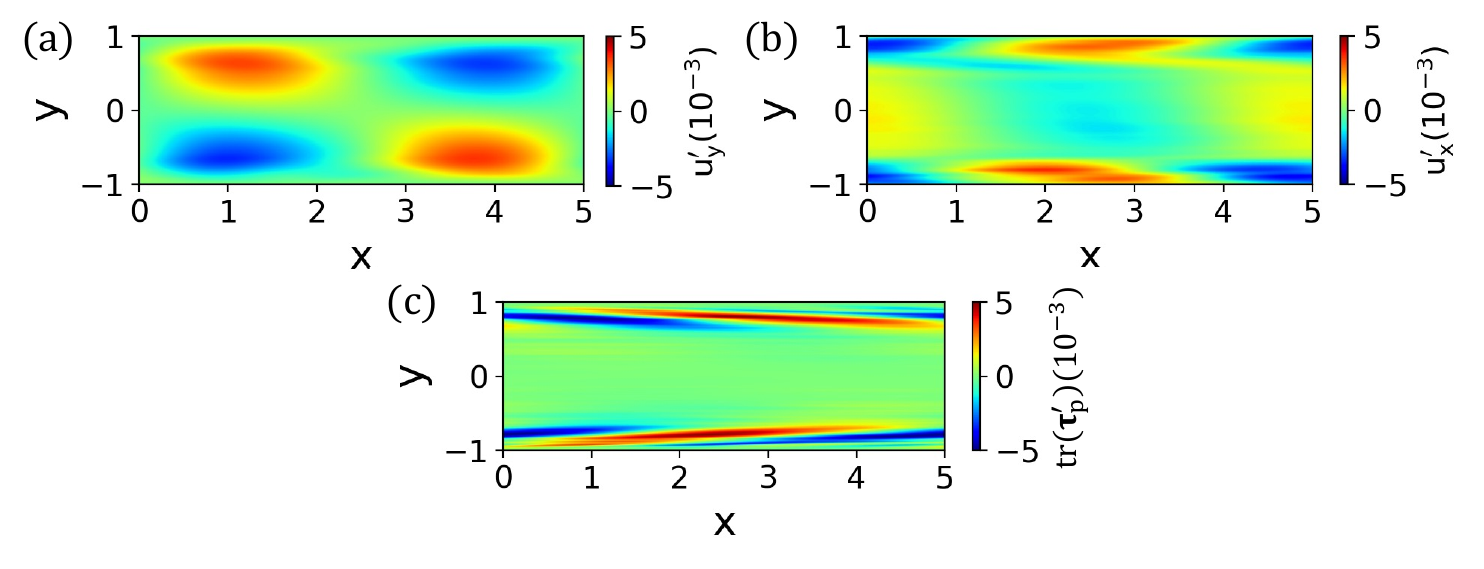}
\caption{Mode structures of the second-most energetic mode at $f=0.08$; (a)  $u_y'$, (b)  $u_x'$, and (c) $\mathrm{tr}(\mathtensor{\tau}_{p}^{\prime})$. Other parameters are $\Rey=3000$ and $\Wi=35$}
\label{uy_thetaxx_m1_f6_Re3000_Wi35.pdf}
\end{figure}

For completeness, we also report in Figure \ref{uy_thetaxx_m1_f6_Re3000_Wi35.pdf}  the structures corresponding to the second-most energetic mode from SPOD (blue curves on Figure \ref{power_specra_Wi35_Re3000_ux_uy_trtau}) at $f=0.08$. Note that the energy of this mode is substantially smaller than that of the leading mode and that the spectrum of this structure does not contain distinct peaks. This mode also has $\kappa=1$ and thus the same wave speed and critical layer position as the most energetic mode. It again exhibits localized polymer stretch fluctuations in the critical layer, but now displays simple reflection symmetry rather than the shift-reflect symmetry of the dominant mode. We view this as a higher-order correction on the dominant structure elucidated above.

\begin{figure}
\centering
\begin{subfigure}[b]{0.48\textwidth} 
\includegraphics[width=\textwidth]{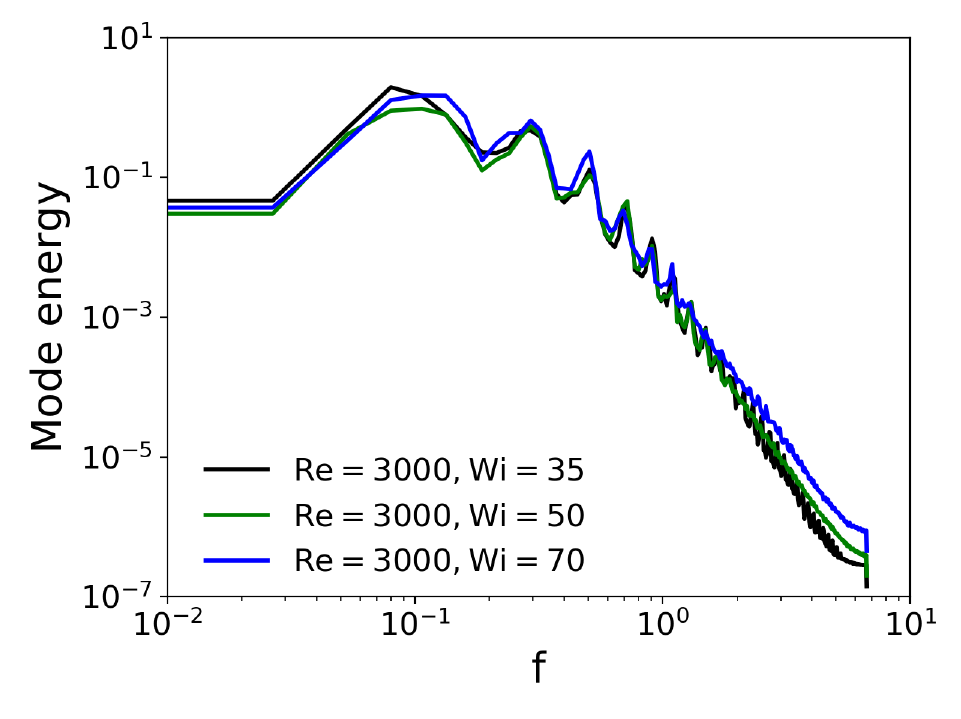}
\caption{}
\label{mode_energy_vs_frequency_uy_Re3000_nfft500_fm8k.png}
\end{subfigure}
\begin{subfigure}[b]{.48\textwidth}
\includegraphics[width=\textwidth]{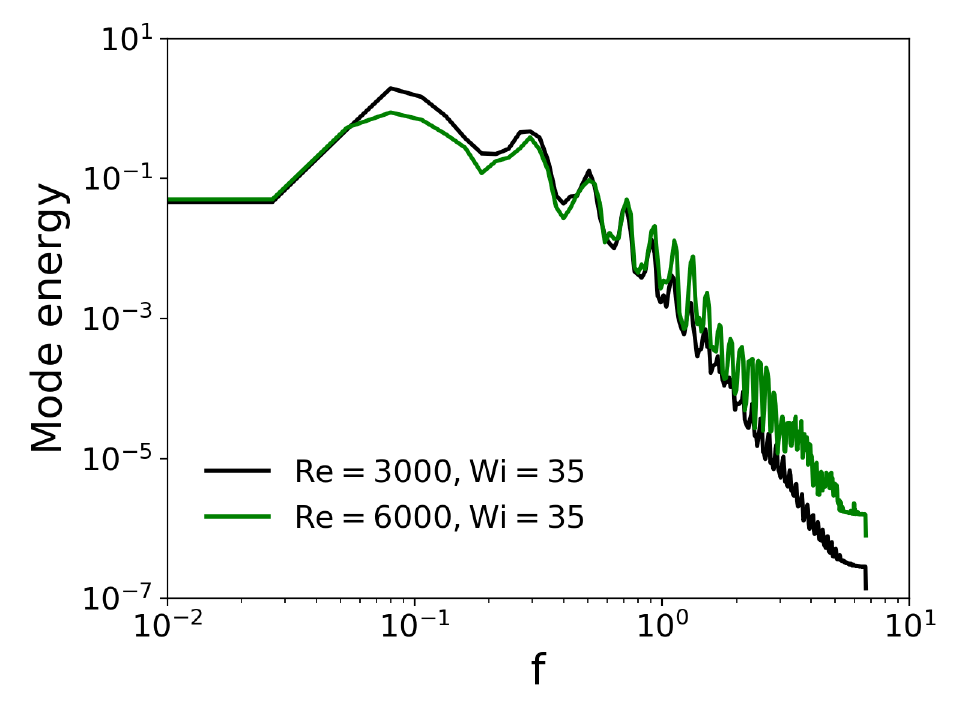}
\caption{}
\label{mode_energy_vs_frequency_uy_Wi35_nfft500_fm8k.png}
\end{subfigure}
\caption{Leading modes of SPOD energy spectra of $u_y'$ at different (a) Wi at $\Rey=3000$ and (b) $\Rey$ at $\Wi=35$.}
\end{figure}

\begin{figure}
\centering
\includegraphics[width=\textwidth]{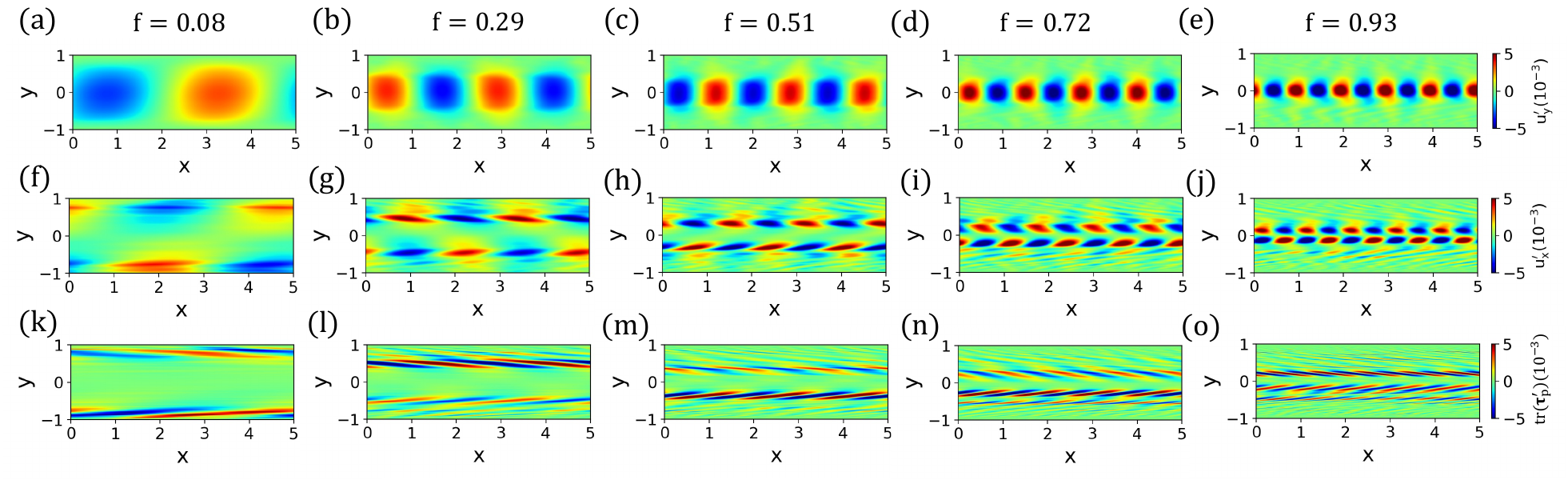}
\caption{Structures of SPOD modes of (a-e) $u_y'$, (f-j) $u_x'$, and (k-o) $\mathrm{tr}(\mathtensor{\tau}_{p}^{\prime})$ at the peak frequencies in the leading mode of $u_y'$ at $\Rey=6000$ and $\Wi=35$.}
\label{uy_trtau_different_frequency_Re6000_Wi35_nfft500.pdf}
\end{figure}

\textcolor{black}{Results obtained for $(\Rey=3000, \Wi=70)$ and $(\Rey=6000, \Wi=35)$ display nearly identical features to the case considered above ($\Rey=3000, \Wi=35$). The leading modes of the SPOD energy spectra of $u_y'$ at different $\Wi$ and $\Rey$ have been shown in Fig. \ref{mode_energy_vs_frequency_uy_Re3000_nfft500_fm8k.png} and Fig. \ref{mode_energy_vs_frequency_uy_Wi35_nfft500_fm8k.png}, respectively. The SPOD spectra of other state variables have peaks exactly at the same frequencies as the spectrum of $u_y'$. Hence, they do not provide any additional information. As $\Wi$ increases, the region close to the first peak in the SPOD spectrum becomes slightly flatter. We do not see any significant effect of $\Wi$ on the peak frequencies in the SPOD spectra, which suggests that $\Wi$ does not have any noticeable impact on the qualitative nature of the traveling structures. As $\Rey$ increases, the mode energy corresponding to the lower wavenumber peaks decreases, whereas the energy corresponding to the higher wavenumber peaks increases. %This suggests that the energy contribution of the higher wavenumber traveling structures in EIT increases with $\Rey$. \MDG{didn't you allready say this or is ther a different point you're trying to make?} 
However, the frequencies corresponding to the peaks in the SPOD spectra remain unchanged indicating that the speed of the traveling wave is independent of $\Rey$. We also plot the SPOD mode structures of $u_y'$, $u_x'$, and $\mathrm{tr}(\mathtensor{\tau}_{p}^{\prime})$ at the frequencies corresponding to the peaks in the leading SPOD mode at $\Rey=6000$ (Fig. \ref{uy_trtau_different_frequency_Re6000_Wi35_nfft500.pdf}). The mode structures of the traveling waves at $\Rey=6000$ are qualitatively similar to the structures at $\Rey=3000$ (Fig. \ref{uy_different_frequency_Re3000_Wi35}).}
% \MDG{we seem to see less shift-reflecy symmetry -- is this significant or related to sampling?} \textcolor{green}{I would not give much emphasis, because velocity field still has pretty good shift-reflect symmetry. The symmetry in the stress field may improve, but due to memory restriction, I can't calculate SPOD for a sample size of more than 8000. It is also possible that at large $\Rey$, EIT is getting more and more complex and hence losing symmetry. }

% \textcolor{green}{I prefer to keep it here as it connects the implication of present results to pipe flows and 3D geometries.}\MDG{ok but you already said most of this in the introduction, some of it verbatim. Shorten substantially to focus on the point you're trying to make.} \textcolor{green}{How is now? If you want I can remove the discussion about center mode from this para.}\MDG{I made some small changes and got rid of the center mode discussion. Put the revised text in the response}
\textcolor{black}{At $\Rey$ relevant to the present study, EIT overwhelmingly represents the MDR state \citep{Lopez:2019ct}. Hence, the self-sustaining chaotic nature of the MDR state and its dynamics can be explained by the EIT. As noted earlier, the dynamics of EIT in both 3D channel and pipe flows are fundamentally two-dimensional \citep{Sid:2018gh, Lopez:2019ct}; specifically, 2D finite amplitude perturbations are self-sustaining. Therefore, it is expected, and observed \citep{Shekar:2019hq}, that EIT even 3D flows would contain traveling waves originating from wall modes similar to the 2D channel flow considered in the present study. At larger $\Rey$, the MDR state may not be fully dominated by 2D EIT, and 3D flow structures may arise in that scenario. An important direction for future work will be to apply SPOD to the full 3D case.} %However, we also note that the stable pure traveling wave solutions in 2D geometry often become unstable in 3D \citep{Lellep2023}. %\MDG{isn't that a zero Reynolds number result -- I think that Ashwin reported the same thing somewhere for EIT}, which may affect the nature of chaotic dynamics and hence the nested traveling wave underlying EIT. \MDG{Leave out -- not the topic of this paper -- we are tryin to reveal the structure of EIT. Making a low-d model is a separate issue.} The existence of a family of nested traveling waves in the chaotic dynamics of EIT can be exploited in the development of a reduced-order model of EIT that will be extremely valuable for understanding, predicting, and even manipulating the dynamic of EIT as a direct numerical simulation of EIT is computationally expensive.

\section{Conclusions}
% polymer additives are commonly used in the pipeline transport of liquids to reduce turbulent drag, which leads to a reduced pumping cost or liquid transfer time. The emergence of elastoinertial turbulence, a chaotic state resulting from the interplay between inertia and elasticity, restricts the turbulent drag reduction due to polymer additives. Dynamics of chaotic states are often organized around well-defined large-scale structures that coherently evolve in time and space and are known as coherent structures. 

In the present study, we use Spectral Proper Orthogonal Decomposition (SPOD) to elucidate the structure underlying the chaotic dynamics of two-dimensional elastoinertial turbulence in channel flow. %The leading SPOD mode contains $75 \%$ and $32 \%$  of total energy in wall-normal velocity and xx-component of conformation tensor, respectively. The reason is that the leading mode mainly captures the structures in the center of the channel and the dynamics of wall-normal velocity in elastoinertial turbulence are dominated by the large-scale structures in the vicinity of the channel center, whereas the polymer stress field is dominated by the fine structures close to the channel walls. 
The most energetic mode of SPOD spectrum has distinct peaks. The mode structures corresponding to these peaks exhibit a family of well-defined traveling structures, where the velocity field contains large-scale regular patterns and the polymer stress field contains the formation of thin inclined sheets of high and low stress at the critical layers of the wave. The structure of the most dominant traveling wave (first mode, highest peak) of this family exhibits shift-reflect symmetry and resembles the structure of the Tollmien–Schlichting wave indicating its origin in a nonlinearly self-sustained wall mode. The traveling structures corresponding to the higher frequency peaks have very similar structure and symmetry, however, their wavenumber increases, and the size of large-scale structures decreases. It appears that the localized polymer sheets at the critical layers of the leading SPOD mode at a given peak frequency act like walls for the traveling structure of the mode corresponding to the next peak, and hence lead to a nested arrangement of the waves. Based on this observation, a simple theory quantitatively captures the relationship between the wave speeds and the locations of critical layers for different waves. From this analysis, a picture emerges of two-dimensional EIT as a nested collection of nonlinearly self-sustaining TS-wave-like structures. %The question of mechanism for this nonlinear self-sustenance remains open. 

%The size of these regular structures decreases and patterns become progressively confined in the vicinity of the channel center at larger frequencies. %The wave speed of these traveling structures initially increases with the frequency and then saturates at a value close to the channel centerline velocity at larger frequencies.The polymer stress field corresponding to these structures contains thin sheets of high and low polymer stresses, which alternate in the streamwise direction and are located at the critical layers of the traveling waves. The critical layers are the positions along the wall-normal direction where wave speed matches the streamwise flow velocity. The structure of the traveling wave at a higher frequency is cocooned inside the wave at a lower frequency, leading to a nested arrangement of the traveling waves underlying elastoinertial turbulence.  %This specific structure in the polymer stress field induces large-scale structures in the velocity field at the channel center. A low-dimensional reconstruction of the velocity field required relatively fewer SPOD modes compared to the polymer stress field to capture a similar proportion of original energy. Because, the dynamics of the velocity field overwhelmingly dominate the channel center and have large-scale compact structures, whereas the dynamics of the stress field are dominated by extremely fine sheets of polymer stress mainly located in the vicinity of channel walls. 

\section*{Acknowledgments}
This research was supported under grant ONR N00014-18-1-2865 (Vannevar Bush Faculty Fellowship). We are grateful to Aaron Towne for helpful discussions and Oliver Schmidt for making available his SPOD code. 

\section*{Declaration of Interests}
The authors report no conflict of interest.

\appendix

%\MDG{can you fix the subsection numbering -- or get rid of it -- we're not in section 4 anymore}
\section{Effect of block size and overlap on SPOD spectra}\label{spod_Nf_No}
\begin{figure}
\centering
\begin{subfigure}[b]{0.32\textwidth} 
\includegraphics[width=\textwidth]{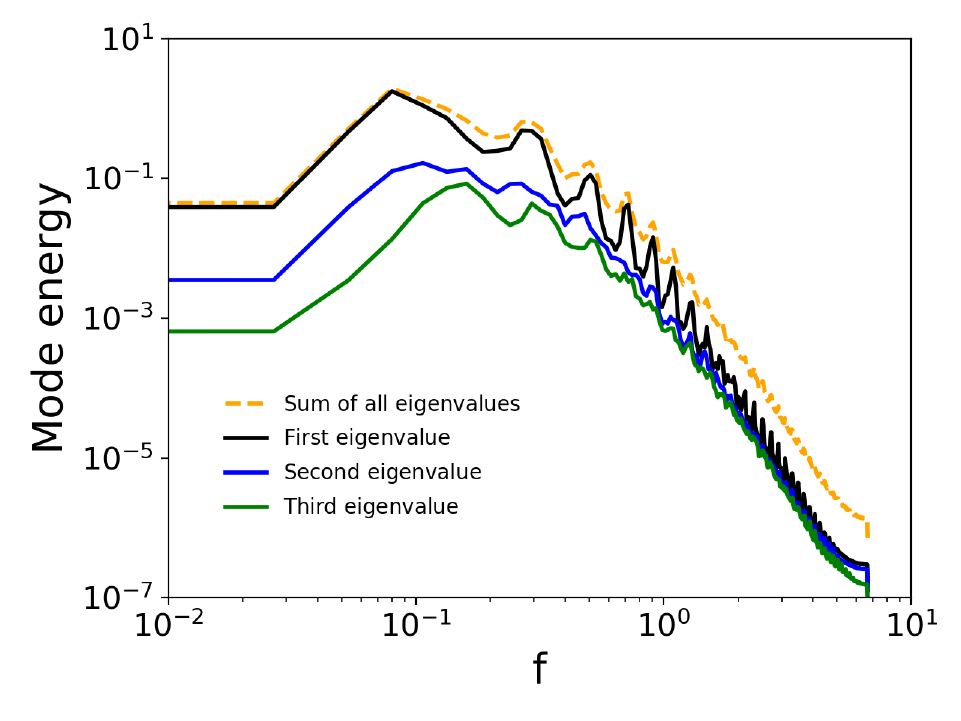}
\caption{}
\label{mode_energy_vs_frequency_uy_Re3000_Wi35_nfft500_fm8k_ovlp125.png}
\end{subfigure}
\begin{subfigure}[b]{.32\textwidth}
\includegraphics[width=\textwidth]{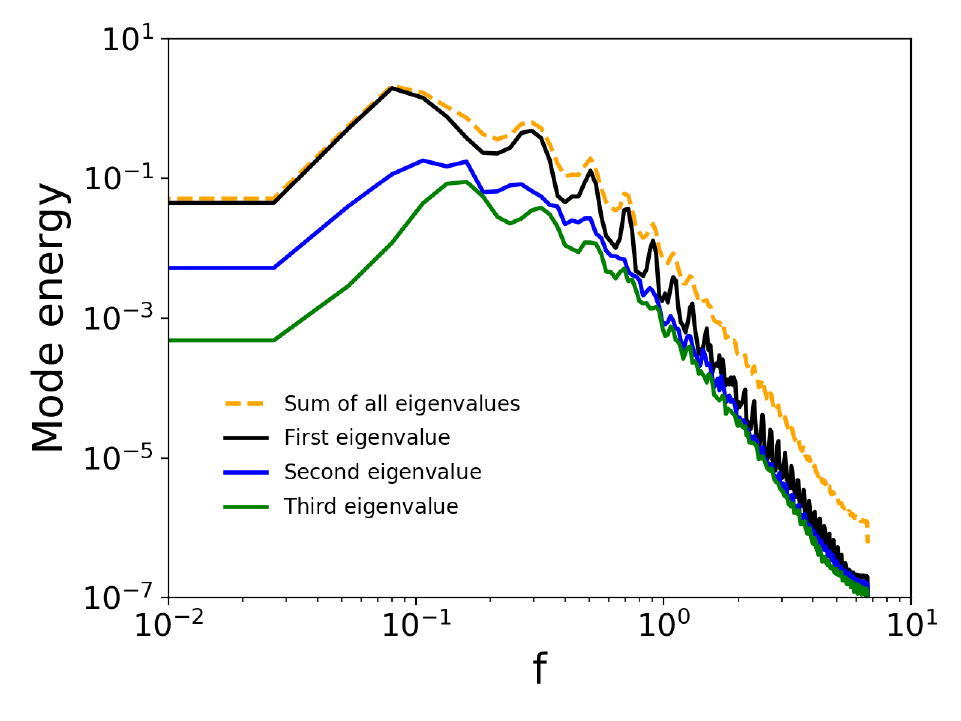}
\caption{}
\label{mode_energy_vs_frequency_uy_Re3000_Wi35_nfft500_fm8k_ovlp375.png}
\end{subfigure}
\begin{subfigure}[b]{.32\textwidth}
\includegraphics[width=\textwidth]{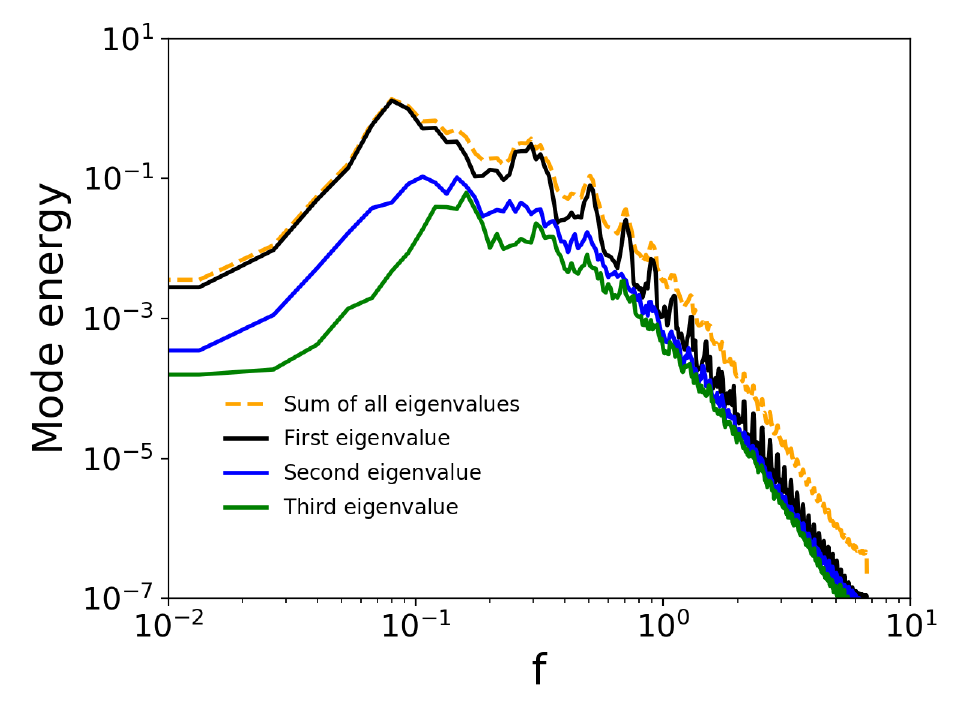}
\caption{}
\label{mode_energy_vs_frequency_uy_Re3000_Wi35_nfft1000_fm8k.png}
\end{subfigure}
\caption{SPOD energy spectra of $u_y'$ at $\Rey=3000$ and $\Wi=35$ estimated using (a) block size $N_f=500$ with $25 \%$ overlap, (b) block size $N_f=500$ with $75 \%$ overlap, and (c) block size $N_f=1000$ with $50 \%$ overlap.} \label{spod_different_block_overlap}
\end{figure}

\textcolor{black}{To investigate the effect of the block size ($N_f$) and overlap on the estimation of SPOD spectra, we plot the SPOD spectra of $u_y'$ obtained using different block sizes and overlaps (Fig. \ref{spod_different_block_overlap}). We do not see any noticeable difference between the peaks in the SPOD spectra obtained using $N_f=500$ with $50 \%$ overlap in the main text (Fig. \ref{power_spectra_uy_Re3000_Wi35.pdf}) and the spectra obtained using different combinations of $N_f$ and overlap (Fig. \ref{spod_different_block_overlap}).} 

\begin{comment}
\section{Mean profile of state variables}\label{mean_profile}
\textcolor{black}{The temporal mean of $u_y'$ is zero (i.e., $\langle u_y \rangle=0$). The temporal mean profiles of streamwise velocity ($\langle u_x \rangle$) and polymer stress ($\langle \mathrm{tr}(\boldsymbol{\tau}_{p}) \rangle$) have been shown in Fig. \ref{mean_trtau_and_uy_vs_y.png}.}  
\begin{figure}
\centering
\includegraphics[width=.5\textwidth]{mean_trtau_and_uy_vs_y.png}
\caption{Mean profile of $u_x$ and $\mathrm{tr}(\boldsymbol{\tau}_{p})$ in EIT at $\Rey=3000$ and $\Wi=35$.\MDG{This should be in the main text, and should also include the laminar profile.  Also looks like you are using angle brackets for multiple purposes here, mean, inner product.}}
\label{mean_trtau_and_uy_vs_y.png}
\end{figure}
\end{comment}

\section{SPOD estimation of velocity components together}\label{SPOD_together}

\begin{figure}
\centering
\includegraphics[width=.5\textwidth]{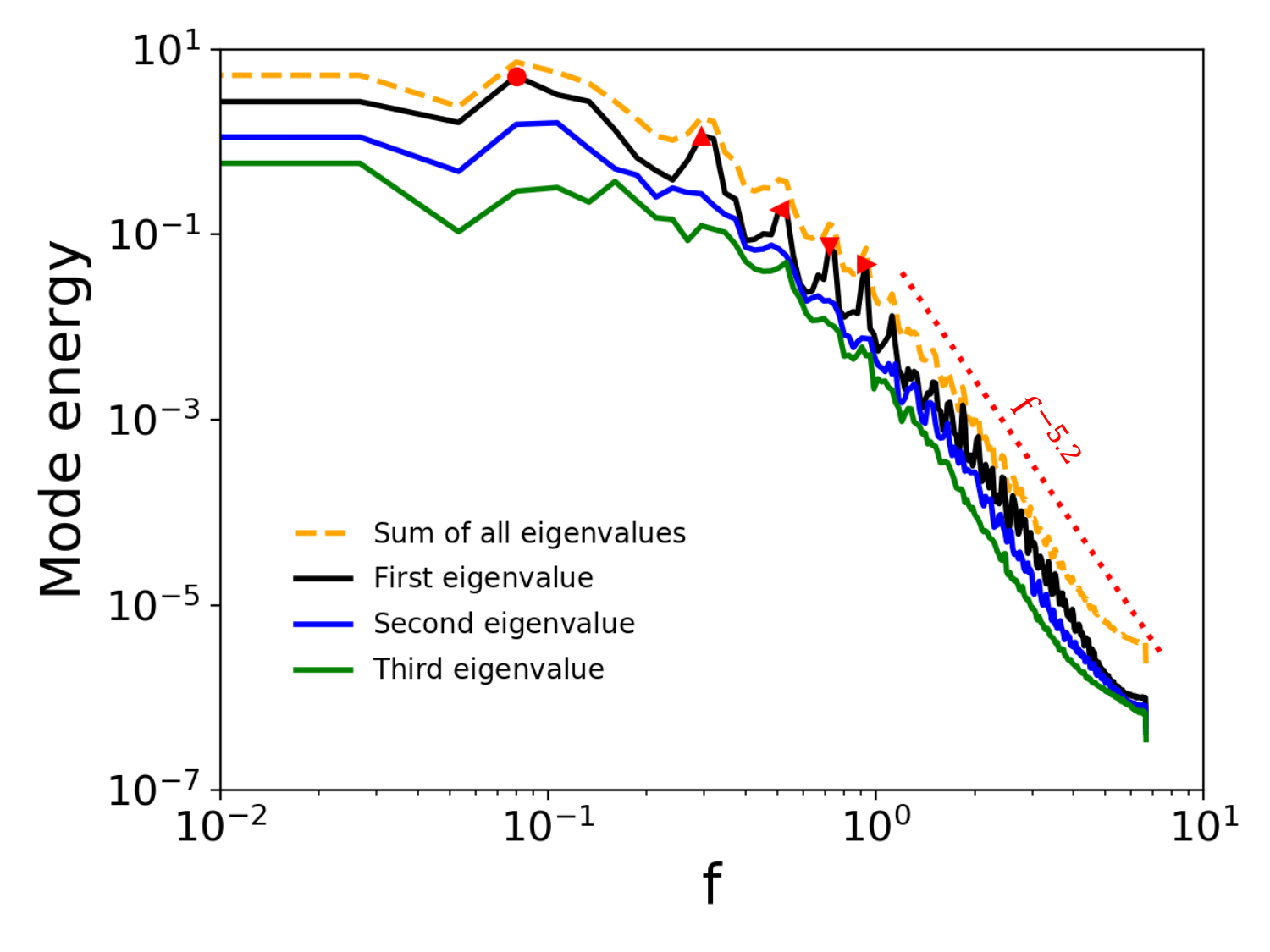}
\caption{SPOD energy spectrum of the velocity field ($u_x'$ and $u_y'$ together) at $\Rey=3000$ and $\Wi=35$ estimated using $N_t=4000$ with $25\%$ overlap.}
\label{power_spectra_uy_ux_Re3000_Wi35.pdf}
\end{figure}

\textcolor{black}{For a small dataset ($N_t=4000$) and lower overlap ($25 \%$), we have calculated the SPOD spectrum of velocity components altogether (Fig. \ref{power_spectra_uy_ux_Re3000_Wi35.pdf}). The peaks in the leading mode are exactly at the same frequencies as the SPOD calculated for each velocity component separately (Figs. \ref{power_spectra_uy_Re3000_Wi35.pdf} and \ref{power_spectra_ux_Re3000_Wi35.pdf}).}  

\section{Quantification of the symmetry of SPOD mode structures}\label{symmetry_quant}
\begin{comment}
To quantify the shift-reflect symmetry of the SPOD mode structures, we define a cross-correlation function as:
\begin{equation}
C=\frac{S(x,y)T(x,y)}{max(S(x,y))max(T(x,y))}, \label{eq:cross_correlation}
\end{equation}
where $S(x,y)$ represents SPOD structures ($u_y'$, $u_x^{\prime}$, and $tr(\boldsymbol{\tau}_{p}^{\prime})$) and $T(x,y)$ represents their shift-reflect image. A large value of $C$ indicates similar structures in both $S$ and $T$ fields (Fig. \ref{cross_correlation_uy_ux_trtau_Re3000_Wi35.pdf}). 
\end{comment}
\textcolor{black}{To quantify the shift-reflect symmetry of the SPOD mode structures, we define a parameter $R$ as: 
\begin{equation}
R=\frac{||A(x,y) - B(x,y)||^2}{4||A(x,y)||^2}, \label{eq:symmetry}
\end{equation}
where $A(x,y)$ represents SPOD structures ($u_y'$, $u_x^{\prime}$, and $\mathrm{tr}(\mathtensor{\tau}_{p}^{\prime})$) and $B(x,y)$ represents their shift-reflect image. $R=0$ and $R=1$ represent perfect shift-reflect symmetry and reflect symmetry, respectively. The small values of $R$ confirm that the SPOD structures obey shift-reflect symmetry (Table \ref{table_sr}). The value of $R$ for $\mathrm{tr}(\mathtensor{\tau}_{p}^{\prime})$ is relatively larger than $u_y'$, because the SPOD modes of $\mathrm{tr}(\mathtensor{\tau}_{p}^{\prime})$ are characterized by thin sheets and even a slight misalignment between sheets leads to a larger value of $R$.} %\MDG{let's do something simpler.  Notice that for shift-reflect symmetry, $u_y-u_y(shift-reflect)=0$ and for reflection symmetry, $u_y-u_y(shift reflect)=2 u_y'$. So we can define a quantity that's zero for shift-reflect and 1 for reflect:  $$||u_y-u_y(shift-reflect||^2/4||u_y||^2$$} \textcolor{green}{I quantify the shift-reflect as you suggested. I prefer to keep both the cross-correlation figure and table 1. R for $u_y'$ is close to zero, but that for stress is bit larger. This is because the stress field has thin sheets and also small-scale fluctuations close to walls. Therefore, R for stress is not as good as that of uy. Should we report R for all the state variables or just uy?}

\begin{table}
\begin{center}
\def~{\hphantom{0}}
  \begin{tabular}{lccccc}
       $f/R$ & $f=0.08$ & $f=0.29$ & $f=0.51$ & $f=0.72$ & $f=0.93$ \\
      $u_y'$ & $0.003$ & $0.009$ & $0.004$ & $0.041$ & $0.158$ \\
      $u_x'$ & $0.058$ & $0.034$ & $0.038$ & $0.211$ & $0.394$ \\
      $\mathrm{tr}(\mathtensor{\tau}_p^{\prime})$ & $0.304$ & $0.089$ & $0.057$ & $0.336$ & $0.397$   \\
  \end{tabular}
  \caption{Values of shift-reflect symmetry value $R$ (Eq.~\ref{eq:symmetry}) for SPOD of various quantities and modes.}
  \label{table_sr}
  \end{center}
\end{table}

\begin{comment}
\begin{figure}
\centering
\includegraphics[width=\textwidth]{cross_correlation_uy_ux_trtau_Re3000_Wi35.pdf}
\caption{Cross-correlation between SPOD mode structures and their shift-reflect images of (a-e) $u_y'$, (f-j) $u_x^{\prime}$, and (k-o) $tr(\boldsymbol{\tau}_{p}^{\prime})$.}
\label{cross_correlation_uy_ux_trtau_Re3000_Wi35.pdf}
\end{figure}
\end{comment}

\section{Reconstruction of polymer stress field using specific SPOD mode} \label{trtaup_reconstruct}
\begin{figure}
\centering
\includegraphics[width=.6\textwidth]{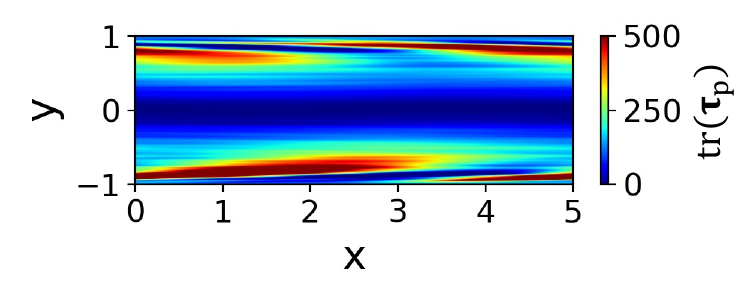}
\caption{Reconstruction of $\mathrm{tr}(\mathtensor{\tau}_{p})$ just using the most dominant SPOD mode (first mode, highest peak) at $\Rey=3000$ and $Wi=35$.}
\label{trtaup_reconstruct_m1_f3.pdf}
\end{figure}

\textcolor{black}{To visualize the signature of the SPOD structure in the polymer stress field, we plot the reconstruction of the polymer stress field just using the most dominant SPOD mode (first mode, highest peak) (Fig. \ref{trtaup_reconstruct_m1_f3.pdf}). In the reconstructed stress field, we see the existence of thin sheets characterized by large polymer stress.
% , which act like barriers for the flow passing through them \citep{Kumar2023stretching} and hence induce nested structures of the traveling wave in EIT.
}

%\MDG{There's still a problem with the Toms reference - His name is not Be A.} \textcolor{green}{I am not sure then what is his first name. This is the name I got from Google Scholar. }\MDG{Then just use B.A. Toms. That's what's given in the original reference. Also, please check capitalization of things like Reynolds in titles}
\bibliographystyle{jfm}
\bibliography{main-nested,turbulence-MDG-2402}

\end{document}